\documentclass[aps,reprint,floatfix,showpacs,superscriptaddress]{revtex4-2}
\usepackage{natbib}
\usepackage{amsfonts,amsmath,amssymb}
\usepackage{bm}
\usepackage{pgf}
\usepackage{dcolumn}
\usepackage{grffile}
\graphicspath{{./plots/}}

\newcommand{\akh}[1]{}

\begin{document}
\title{The Griffiths phase and beyond: A large deviations study of the magnetic susceptibility of the two-dimensional bond-diluted Ising model}
\author{L. M\"unster}
\affiliation{Institut f\"ur Physik, Technische Universit\"at Chemnitz, 09107 Chemnitz, Germany}
\author{A. K. Hartmann}
\affiliation{Institut f\"ur Physik, Universit\"at Oldenburg, 26111 Oldenburg, Germany}
\author{M. Weigel}
\affiliation{Institut f\"ur Physik, Technische Universit\"at Chemnitz, 09107 Chemnitz, Germany}
\date{\today}

\begin{abstract}
  The Griffiths phase in systems with quenched disorder occurs below the ordering
  transition of the pure system down to the ordering transition of the actual
  disordered system. While it does not exhibit long-range order, large fluctuations
  in the disorder degrees of freedom result in exponentially rare, long-range ordered
  states and hence the occurrence of broad distributions in response
  functions. Inside the Griffiths phase of the two-dimensional bond-diluted Ising
  model the distribution of the magnetic susceptibility is expected to have such a
  broad, exponential tail. A large-deviations Monte Carlo algorithm is used to
  sample this distribution and the exponential tail is extracted over a wide range of
  the support down to very small probabilities of the order of $10^{-300}$. We study
  the behavior of the susceptibility distribution across the full phase diagram, from
  the paramagnetic state through the Griffiths phase to the ferromagnetically ordered
  system and down to the zero-temperature point. We extract the rate function of
  large-deviation theory as well as its finite-size scaling behavior and we reveal
  interesting differences and similarities between the cases. A connection between
  the fraction of ferromagnetic bonds in a given disorder sample and the size of the
  magnetic susceptibility is demonstrated numerically.
\end{abstract}
\maketitle
\section{Introduction}\label{sec:introduction}

The Ising model is one of the most studied systems in statistical physics and has
found applications in many branches of science, see, e.g.,
Refs.~\cite{Brush1967HistoryOfTheLenzIsingModel,ConiglioFierro2021CorrelatedPercolation,AmandaSear2006HeterogeneousNucleationInAndOutOfPores,Grabowski2006IsingBasedModelOfOpinionFormationInAComplexNetworkOfInterpersonalInteractions,Sethna2001CracklingNoise,NewmanStein2013SpinGlassesAndComplexity,Nishimori2001StatisticalPhysicsOfSpinGlassesAndInformationProcessing}.
Its original purpose is to describe ferromagnetism in homogeneous materials with
strong uniaxial asymmetry. Often, however, materials are not perfectly homogeneous
but exhibit randomly distributed impurities. In such cases, a central question is
how these random impurities affect the physical properties of the system in
comparison to the idealized pure model~\cite{NishimoriOrtiz2011RandomSystems,Dotsenko1983CriticalBehaviourOfThePhaseTransitionInThe2DIsingModelWithImpurities,Shalaev1994CriticalBehaviorOfThe2DIsingModelWithRandomBonds}. Impurities can be incorporated into the Ising model, for instance, by randomly
removing a fraction $1-p$ of the bonds that represent the ferromagnetic interactions
between the spins, where $0 \leq p \leq 1$. Due to the overall weakened ferromagnetic
coupling this leads to a shift in the transition temperature from
$T_\mathrm{f}:=T_\mathrm{c}(p=1)$ in the pure system to a lower temperature
$T_\mathrm{c}(p),~p<1$ in the diluted system. As shown by
Harris~\cite{ABHarris1974EffectOfRandomDefectsOnTheCriticalBehaviourOfIsingModels},
the critical behavior at the transition differs from that found in the pure system if
the specific heat exponent $\alpha$ of the pure system is positive. This is the case
for the Ising model in dimensions $d \ge 3$, while $\alpha = 0$ in $d=2$. Due to this
marginality one expects logarithmic corrections to the leading critical behavior in
two dimensions~\cite{LudwigCardy1987PerturbativeEvaluationOfTheConformalAnomalyAtNewCriticalPointsWithApplicationsToRandomSystems,Ludwig1987CriticalBehaviorOfThe2DRandomQStatePottsModelByExpansionInQMinus2,Shankar1987ExactCriticalBehaviorOfArandomBondTwoDimensionalIsingModel,Ludwig1988CommentOnShankar1987,Shalaev1994CriticalBehaviorOfThe2DIsingModelWithRandomBonds}, but according to numerical
simulations~\cite{HasenbuschEtAl2008UniversalDependenceOnDisorderOf2DRandomlyDilutedAndRandomBondPMJIsingModels,MartinsPlascak2007UniversalityClassOfThe2DSiteDilutedIsingModel,WangEtAl1990TheCriticalBehaviourOfThe2DDiluteIsingMagnet,AndreichenkoEtAl1990MonteCarloStudyOfThe2DIsingModelWithImpurities,Hadjiagapiou2011MonteCarloAnalysisOfTheCriticalPropertiesOfThe2DRBDIMviaWangLandauAlgorithm}
as well as experiments~\cite{IdekaEtAl1979NeutronScatteringInvestigationOfStaticCriticalPhenomenaInTheTwoDimensionalAntiferromagnets,Ferreira1983RandomFieldInducedDestructionOfThePhaseTransitionOfADiluted2DIsingAntiferromagnet} the critical exponents remain identical to the pure case.

The thermal region between $T_\mathrm{f}$ and $T_\mathrm{c}(p<1)$ is known as the
\emph{Griffiths phase}~\cite{Griffiths1969NonanalyticBehaviorAboveTheCriticalPointInARandomIsingFerromagnet}. In this regime the order parameter of the ferromagnetic phase transition, the
magnetization, remains zero, but arbitrarily large fluctuations of the order
parameter become possible. These fluctuations can exist due to large compact
structures of ferromagnetic bonds in regions where there are less missing bonds in
comparison to the overall average of the system. Within these structures the system
is effectively in a ferromagnetic state such that a change of the spin orientation
happens coherently, giving rise to large fluctuations in the magnetization. As a
consequence the distribution of the second moment of the magnetization, the magnetic
susceptibility, is expected to have an exponential tail that extends to infinity,
which is an expression of the essential, but weak, Griffiths
singularity~\cite{Bray1987NatureOfTheGriffithsPhase,BrayMoore1982OnTheEigenvalueSpectrumOfTheSusceptibilityMatrixForRandomSpinSystems}:
Since the magnetic susceptibility characterizes the response of the system to an
external magnetic field, the free energy is a non-analytic function of the field
throughout the Griffiths
phase~\cite{Griffiths1969NonanalyticBehaviorAboveTheCriticalPointInARandomIsingFerromagnet}.
Besides these effects on static averages, the Griffiths singularity also plays an
important role for the dynamics of the system, leading to a slow-down of the decay of
the spin-spin correlation
function~\cite{Bray1988DynamicsOfDiluteMagnetsAboveTc,Bray1987NatureOfTheGriffithsPhase,ColborneBray1989MonteCarloStudyOfGriffithsPhaseDynamicsInDiluteFerromagnets,
  HengEtAl2006MonteCarloStudyOfGriffithsPhaseInRandomlySiteDilutedIsingMagneticSystem}. The Griffiths singularity does not only occur in the diluted Ising ferromagnet~\cite{BrayHuifang1989GriffithsSingularitiesInRandomMagnetsResultsForASolubleModel}
but it may also be observed in other disordered systems such as spin
glasses~\cite{MatsudaEtAl2008TheDistributionOfLeeYangZerosAndGriffithsSingularitiesInThePMJModelOfSGs}. The
analogous quantum mechanical effect is known as the Griffiths-McCoy singularity, and
it has been studied theoretically, numerically as well as in
experiments~\cite{Vojta2010QuantumGriffithsEffectsAndSmearedPhaseTransitionsInMetalsTheoryAndExperiment,Fisher1992RandomTransverseFieldIsingSpinChains,Fisher1995CriticalBehaviorOfRandomTransverseFieldIsingSpinChains,YoungRieger1996NumericalStudyOfTheRandomTransverseFieldIsingSpinChain,
  PichEtAl1998CriticalBehaviorAndGriffithsMcCoySingularitiesInThe2DRandomQuantumIsingFerromagnet,WangEtAl2017QuantumGriffithsPhaseExperiment,NishimuraEtAl2020GriffithsMcCoySingularityonTheDilutedChimeraGraph}.

In the present work we use numerical simulations \cite{practical_guide2015} based on
the Monte Carlo method \cite{NewmanBarkema1999MonteCarloMethodsInStatisticalPhysics}
to explore the Griffiths phase by investigating the distribution over the bond
disorder of the magnetic susceptibility in the two-dimensional bond-diluted Ising
ferromagnet. In particular, we are interested in the tail of the distribution. To
obtain this tail we employ a large-deviation sampling
algorithm~\cite{Hartmann2002SamplingRareEventsStatisticsOfLocalSequenceAlignments}
which has previously proved useful for a variety of rare-event sampling problems. A
drawback of the variant of the algorithm used to date is that for some cases it does
not scale well with increasing system
size~\cite{Hartmann2011LargeDeviationPropertiesOfLargestComponentForRandomGraphs}.
For the system at hand, we resolve this problem by using a different bias for the
sampling process, see Sec.~\ref{sec:large_deviations_sampling} for details. The
authors of a previous study of the bond-diluted Ising
model~\cite{HukushimaIba2008AMonteCarloAlgorithmForSamplingRareEvents} performed a
similar analysis of the magnetic susceptibility, but they were only able to sample
relatively closely to the mean value of the distribution. Here, the distribution
will be presented over a wide range of its support. Furthermore, the distribution of
the magnetic susceptibility is also studied at the critical temperature and inside
the ferromagnetic phase. Interestingly, an exponential tail is also found for the
distribution inside the ferromagnetic phase but the mechanism leading to it appears
to be different from that in the Griffiths phase.

The rest of this paper is organized as follows. In Sec.~\ref{sec:bdif_model} we
introduce the two-dimensional bond-diluted Ising model and discuss its essential
properties in so far as they are relevant in the context of the present study. In
Sec.~\ref{sec:large_deviations_sampling} the large-deviation sampling algorithm of
Ref.~\cite{Hartmann2002SamplingRareEventsStatisticsOfLocalSequenceAlignments} is
summarized and we introduce a weight-construction scheme based on ideas of Neuhaus
and
Hager~\cite{NeuhausHager2006FreeEnergyCalculationsWithMultipleGaussianModifiedEnsembles}. In
Sec.~\ref{sec:results_griffiths} we present our simulation results for the disorder
distribution of the magnetic susceptibility inside the Griffiths phase and at the
critical temperature, while Sec.~\ref{sec:results_below_tc} is devoted to our results
for the distribution inside the ferromagnetic phase and at zero temperature. Finally,
Sec.~\ref{sec:discussion} contains a discussion and outlook.

\section{The Two-Dimensional Bond-Diluted Ising Ferromagnet}
\label{sec:bdif_model}

\begin{figure}
  \begin{center}
    \includegraphics{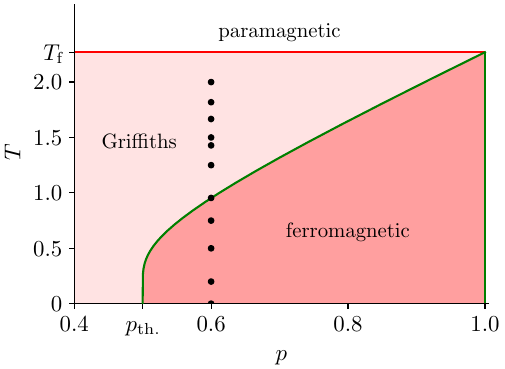}
    \caption{Zero-field phase diagram of the two-dimensional
      bond-diluted Ising ferromagnet as a function of the fraction $p$ of
      ferromagnetic bonds and the temperature $T$.  The boundary between
      the ferromagnetic and paramagnetic phase is obtained according to
      the ``$s=1$'' technique of
      Ohzeki~\cite{Ohzeki2009LocationsOfMulticriticalPointsForSpinGlassesOnRegularLattices}.
      The black dots at $p=0.6$ mark the temperatures at which our
      simulations are performed.  }
    \label{fig:phase_diagram}	
  \end{center}	
\end{figure}

The Hamiltonian of the bond-diluted Ising ferromagnet is given by  
\begin{align}
	\hat{H}_{\bm{J}}\left( \bm{S} \right)=-\sum_{\langle \bm{x}, \bm{y} \rangle} J_{\bm{x}\bm{y}} s_{\bm{x}} s_{\bm{y}}-h\hat{M},
\end{align}
where $\bm{S} \in \{-1,1\}^N$ denotes a spin configuration, and
$\hat{M}=\sum_{\bm{x}}s_{\bm{x}}$ represents the magnetization which
couples to the external magnetic field $h$. The Ising spins $s_{\bm{x}}=\pm 1$ are
placed at the sites $\bm{x}$ of a two-dimensional square lattice of linear dimension
$L$, resulting in $N=L^2$ spins in total. The notation
$\langle \bm{x}, \bm{y} \rangle$ refers to summation over nearest neighbors only. The
time-independent, quenched interaction between two spins is represented by the
exchange coupling $J_{\bm{x}\bm{y}}$. To incorporate random dilution into the model,
the bonds are drawn from a bimodal distribution, such that the probability to obtain
a particular bond sample $\bm{J} = \{ J_{\bm{xy}} \}$ is given by
\begin{align}
 P_J(\bm{J}) = \prod_{\langle \bm{x}, \bm{y} \rangle} p\,  
   \delta[J_{\bm{xy}}-1] +
  (1-p)\delta[J_{\bm{xy}}],
  \label{eq:bond_ensemble}
\end{align} 
where we use $\delta[x]$, $x\in \mathbb{R}$, as an indicator function that yields one
if $x=0$ and zero otherwise. Here, $p$ corresponds to the probability of drawing a
ferromagnetic bond with $J_{\bm{x}\bm{y}}=1$, while $1-p$ is the probability of
missing bonds with $J_{\bm{x}\bm{y}}=0$. The model is studied in the canonical
ensemble at temperature $T$, such that the spin configurations $\bm{S}$ are Gibbs-Boltzmann
distributed according to
\begin{align}
  P_S(\bm{S}| \bm{J})=\frac{1}{Z_{\bm{J}}} \exp\left \{ -\hat{H}_{\bm{J}}(\bm{S})/T\right \},
  \label{eq:spin_ensemble}
\end{align}
where we have set the Boltzmann constant $k_\mathrm{B} := 1$ for convenience.
$Z_{\bm{J}}$ is the partition function for a given bond sample $\bm{J}$.  The thermodynamic state of the model in the absence of an external magnetic field, $h=0$, depends on
two parameters, the bond occupation probability $p$ and the temperature $T$. The
order parameter which can be used to determine the phase of the thermodynamic state is the
first moment of the magnetization per site \footnote{Note that below the
  ferromagnetic phase transition the symmetry is broken such that there are two thermodynamic states, one with positive and one with negative magnetization. To obtain the
  correct value for the magnetization, the average has to be restricted to one of the states. In numerical simulations, however, there is usually no symmetry
  breaking and thus one often uses the absolute value of the magnetization as an
  estimator for the magnetization or, alternatively, the largest FKCK cluster, see
  Sec.~\ref{sec:large_deviations_sampling}.}
\begin{align}
	\overline{m} =\left [  \left \langle \hat{m} \right \rangle_S \right]_J,
\end{align}
where $\hat{m}=\hat{M}/N$. Here, $\langle \cdot \rangle_S$ denotes the thermal
average with respect to the Gibbs-Boltzmann distribution, Eq.~\eqref{eq:spin_ensemble}, and
$[\cdot]_J$ is the average with respect to the disorder distribution,
Eq.~\eqref{eq:bond_ensemble}. As is clear from the phase diagram of the system shown
in Fig.~\ref{fig:phase_diagram}, there exists a high temperature paramagnetic phase
and a low temperature ferromagnetic phase. The boundary between these phases extends
from the bond-percolation threshold $p_{\mathrm{th}}=0.5$ at which the transition
temperature is zero, $T_\mathrm{c}(0.5)=0$, to that of the pure ferromagnet with
$p = 1$ and $T_\mathrm{f}=T_\mathrm{c}(1)=2/\ln(1+\sqrt{2})= 2.2691 \dots $. In
between these limits, the phase boundary can be obtained by arguments based on
duality~\cite{Nishimori1979ConjectureOnTheExactTransitionPointOfTheRandomIsingFerromagnet,Ohzeki2009LocationsOfMulticriticalPointsForSpinGlassesOnRegularLattices}
which yield a good estimate for the true curve, consistent with what is found in
numerical simulations~\cite{ZhongEtAl2020SuperSlowingDownInTheBondDilutedIsingModel}.
The thermal region between the ferromagnetic phase transition of the pure system at
$T_\mathrm{f}=T_\mathrm{c}(1)$ and the ferromagnetic phase transition in the diluted
system $T_\mathrm{c}(p<1)$ is known as the Griffiths phase. Inside the Griffiths
phase the order parameter remains zero but large fluctuations of the magnetization
are more likely than in the paramagnetic phase. These fluctuations are visible in the
magnetic susceptibility $\chi_{\bm{J}}$ which can be defined from the variance of the
magnetization per lattice site for a given bond sample $\bm{J}$,
\begin{align}
  \chi_{\bm{J}} = N 
  \left( \left \langle  \hat{m} ^2 \right \rangle_S -\langle \hat{m}\rangle_S^2 \right) ,
  \label{eq:magnetic_magnetic susceptibility}
\end{align} 
where $\langle \hat{m} \rangle_S=0$ if $T\geq
T_\mathrm{c}$. Bray~\cite{Bray1987NatureOfTheGriffithsPhase} has predicted that the
probability distribution of the magnetic susceptibility over the bond samples
displays an exponential tail throughout the Griffiths phase. The functional form of
this tail was derived to follow the form~\cite{Bray1987NatureOfTheGriffithsPhase}
\begin{align}
  P_{\chi}\left(\chi \right)\sim   \exp \left( - A \chi - 2 \ln{\chi}  \right)
  \label{eq:exponential_tail_bray}
\end{align}
for $\chi \to \infty$. Here, $A$ is a temperature-dependent positive constant which
vanishes at the ferromagnetic phase transition (implying a particularly broad
distribution there) and diverges when $T$ approaches $T_\mathrm{f}$ (such that the
tail disappears). Since the derivation which leads to
Eq.~\eqref{eq:exponential_tail_bray} includes variational arguments and a number of
approximations, Bray concluded that this form may only constitute a lower bound for
the true tail. The magnetic susceptibility describes the linear response of the
magnetization to an external magnetic field, since
$(\partial/\partial h) \langle \hat{m} \rangle_S = \chi_{\bm{J}}/T$. Correspondingly,
inside the Griffiths phase the free energy is a non-analytic function of the external
magnetic
field~\cite{Griffiths1969NonanalyticBehaviorAboveTheCriticalPointInARandomIsingFerromagnet}.

The origin of the large values of the magnetic susceptibility inside the Griffiths
phase are compact structures of ferromagnetic bonds. In these local structures the
fraction of ferromagnetic bonds is larger than the average expected fraction $p$ of
the infinite system. Below the ferromagnetic phase transition, i.e., for
$T<T_\mathrm{c}(p<1)$, there also can be large fluctuations in the magnetization, but
these are caused by a different mechanism which we will explore below. In both cases,
bond samples which lead to larger than average values of the magnetic susceptibility
occur rather rarely. Therefore, to numerically investigate the disorder distribution
of the magnetic susceptibility it is necessary to employ large-deviation sampling
techniques. These are the subject of the next section.

\section{Large Deviations Sampling}
\label{sec:large_deviations_sampling}

To sample the distribution of the magnetic susceptibility over a wide range of the
support we use the large-deviation Monte Carlo algorithm proposed in
Ref.~\cite{Hartmann2002SamplingRareEventsStatisticsOfLocalSequenceAlignments}. The
basic idea of this method is to utilize an auxiliary Markov chain Monte Carlo process
in the disorder degrees of freedom that is biased in such a way that it creates bond
configurations that lead to the desired values in a quantity of interest such as, in
our case, the susceptibility. The bias is then removed \emph{a posteriori} by
reweighting, such that at the end the actual distribution of interest is obtained.

To be more specific, assume that the disorder dependent quantity of interest is
$Y_{\bm{J}}$, where in our particular case $Y_{\bm{J}}=\chi_{\bm{J}}$ and thus
$Y_{\bm{J}}\geq 0$ \footnote{Note that in most cases we used $Y_{\bm{J}}=\sqrt{N \chi_{\bm{J}}}$ instead of
  $\chi_{\bm{J}}$ since this quantity is found to be easier to sample. The
  distribution of $\chi_{\bm{J}}$ can then be obtained by a change of variables.}. The
probability distribution of this quantity can be written as
\begin{align}
     P_Y(Y) = \sum_{\bm{J}} P_{J}(\bm{J}) \delta[Y_{\bm{J}}-Y]
\end{align} 
where $P_J(\bm{J})$ is the unbiased bond distribution as given in
Eq.~\eqref{eq:bond_ensemble}.  If we draw bonds from the unbiased distribution, this
will only give us the typical values of $Y_{\bm{J}}$ in the region where the
distribution $P_Y(Y)$ has most of its weight.  To receive bond samples which lead to
$Y_{\bm{J}}$ in the range of interest, i.e., where $P_Y(Y)$ is extremely small, we
introduce a Monte Carlo process that generates samples from a biased bond
distribution,
\begin{align}
	\widetilde{P}_J(\bm{J};\Theta)=  \frac{P_J(\bm{J})f_{\Theta}(Y_{\bm{J}})}{Z_{\Theta}} . 
	\label{eq:biased_bond_ensemble}
\end{align}  
Here, $Z_{\Theta}$ is a normalization constant that will be determined later, and
$f_\Theta(Y_{\bm{J}})$ is a bias function which depends on the set of parameters
$\Theta$. The bias function has to be chosen such that $\widetilde{P}_J(\bm{J};\Theta)$
defines a distribution which has enough weight in the regions that one would like to
sample. Following the proposal of Neuhaus and Hager in
Ref.~\cite{NeuhausHager2006FreeEnergyCalculationsWithMultipleGaussianModifiedEnsembles} to use a sequence of Gaussians centred at successive values of the reaction coordinate in order to bias the probability density in a region of suppressed weight, we propose to use a generalized exponential of
the form
\begin{align}
	f_{\Theta}\left(Y_{\bm{J}} \right)=\exp\left \{ \theta_1 Y_{\bm{J}}-\frac{\theta_2^2}{2}\left( Y_{\bm{J}} - \theta_3  \right )^2 \right \}
\end{align} 
with $\Theta=(\theta_1,\theta_2,\theta_3)\in\mathbb{R}^3$ to bias the sampling for the purpose of accessing rare events.
To understand the effect of this bias in detail, at first set $\theta_2=0$ such
that $f_\Theta$ becomes a simple exponential. Now, if $\theta_1<0$ the weight of the
bond samples with small $Y_{\bm{J}}$ in comparison to typical values of the unbiased
distribution increases, and if $\theta_1>0$ bond samples with large $Y_{\bm{J}}$ have
more weight. If, on the other hand, we set $\theta_1$ to zero, $f_\Theta$ has the
shape of a Gaussian of width $2/\theta_2$ centered around $\theta_3$. This means that the
bias function will increase the weight close to $\theta_3$. With an adequate
combination of the parameters in $\Theta$ it is possible to create a biased bond
distribution with enough weight in regions that lead to the desired values of
$Y_{\bm{J}}$. While this recipe is fairly general, we do not exclude the possibility
that for some problems other types of bias functions might work better.

\begin{table}
\renewcommand*{\arraystretch}{1.2}
    \begin{tabular}{|c||c|c|c|c|c|c|c|c|c|c|c|}		
			\hline
			& \multicolumn{11}{c|}{$\Theta_k,~k=1,\dots,11$} \\
			\hline
			$~k~~$ & ~1~ & ~2~ &~3~ &~4 ~&~ 5~ &~ 6~ & ~7 ~& ~8~ & ~9~ & ~10~ &~ 11~ \\
			\hline
			$~\theta_1~$ & -3 & 0 &0.9 &1 & 1 & 1 & 1 & 1 & 1& 1.5 & 3 \\
			\hline
			$~\theta_2~$ & 0 & 0 & 0 & 1/5 & 1/5 & 1/5 & 1/5 & 1/5 & 1/5& 0 & 0 \\	
			\hline
			$~\theta_3~$ & 0 & 0& 0 & 10 & 20 & 30 & 40 & 50 & 60& 0 & 0\\
			\hline
		\end{tabular}
                \caption{Simulation parameters used to generate the histogram data
                  shown in Fig.~\ref{fig:ysampling}(b). There are eleven parameter
                  sets in total, $\Theta_k,~k=1,\dots,11$. The parameter set $k=2$
                  corresponds to unbiased sampling. For $k=3$ the Monte Carlo process
                  oscillates between small and large values.}
  \label{tab:ysampling}
\end{table}

To sample from the biased bond distribution $\widetilde{P}_J$, the Metropolis-Hastings
algorithm is used, which in this particular case works as follows. Suppose that we
start with a bond sample $\bm{J}_\mu$ with the corresponding value $Y_{\bm{J}_\mu}$
of the observable of interest. Now, one generates a candidate bond sample
$\bm{J}_{\mu}'$ by randomly selecting one or more bonds of $\bm{J}_\mu$ and assigning
to them new coupling values according to the unbiased distribution
Eq.~\eqref{eq:bond_ensemble}. In other words, a randomly selected bond is set to $1$ with
probability $p$ and to $0$ with probability $1-p$. After computing the observable
value for the proposed sample, $Y_{\bm{J}_{\mu}'}$, the new bond configuration is
accepted with probability
\begin{align}
  A\left( \bm{J}_{\mu} \to \bm{J}_{\mu}' \right) &=
                                                   \mathrm{min} \left \{ 1, \frac{f_\Theta \left(Y_{\bm{J}_{\mu}'} \right)}{f_\Theta \left( Y_{\bm{J}_{\mu}^{~} } \right)} \right \}. 
  \label{eq:bond_metropolis}
\end{align} 
As a result, the new sample in the Monte Carlo chain $\bm{J}_{\nu}$ will be
$\bm{J}_{\nu}=\bm{J}_{\mu}'$ if the proposed sample is accepted, or
$\bm{J}_{\nu}=\bm{J}_{\mu}$ otherwise.

\begin{figure}[tb!]
  \begin{center}
    \includegraphics{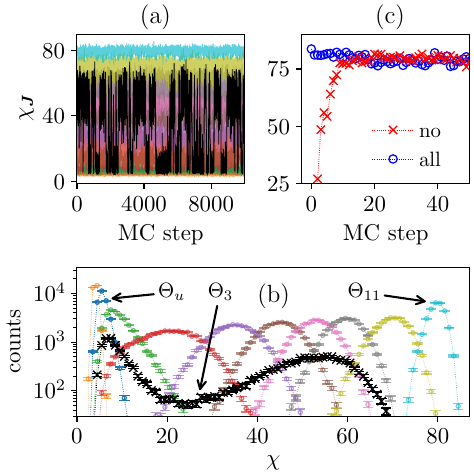}	
    \caption{
      Sampling procedure of the large-deviation algorithm to
      sample the magnetic susceptibility of the two-dimensional
      bond-diluted Ising ferromagnet at $T=2$ with $L=10$ and
      $p=0.6$. (a) Estimates of $Y_{\bm{J}}=\chi_{\bm{J}}$ in the
      bond Monte Carlo chain generated by the biased bond
      distribution, see Eq.~\eqref{eq:biased_bond_ensemble}, for various
      sets of parameters. (b) Histograms of sampled observable values in
      the different biased ensembles after a binning of the data from the
      different Monte Carlo chains. The corresponding parameters are
      listed in Table \ref{tab:ysampling}. The unbiased distribution
      corresponds to $\Theta_u=\Theta_2=(0,0,0)$. The set
      $\Theta_3=(0.9,0,0)$ is special because in this case the observable
      oscillates between small and large values. The corresponding data
      in (a) and (b) is colored black. (c) Equilibration phase of a Monte
      Carlo process for $\Theta_{11}=(3,0,0)$. The Monte Carlo chains are
      initialized with no bonds or all bonds present, respectively. After
      approximately $20$ Monte Carlo steps both chains start to oscillate
      around the same value indicating that equilibrium has been
      reached.}
\label{fig:ysampling}
	\end{center}
\end{figure}

Finally, we establish a connection to the unbiased bond distribution by noticing that
\begin{align*}
\widetilde{P}_Y \left( Y; \Theta \right)  &= \sum_{\bm{J}} \widetilde{P}_J\left( \bm{J}; \Theta  \right) \delta \left[Y_{\bm{J}} - Y   \right] \\  
   & = \frac{f_{\Theta}(Y)}{Z_\Theta}P_Y\left( Y \right)
\end{align*}
and hence
\begin{align}
	P_Y(Y)  = \frac{Z_\Theta}{ f_{\Theta}(Y)} \widetilde{P}_Y\left( Y; \Theta \right). 
	\label{eq:bias_correct}
\end{align}
If one would like to sample $P_Y$ over a wide range of the support, it is possible to
use multiple parameter sets $\Theta_k$, $k=1$, $2$, $\dots$, $K$. After completing
the simulations, the biases are corrected by utilizing Eq.~\eqref{eq:bias_correct}.
In order to achieve this, it is necessary to determine the constants $Z_{\Theta_k}$
which can be deduced from the continuity of the overall distribution. In other words,
in regions where two distributions with parameters $\Theta_i$ and $\Theta_j$ overlap,
their weight should be identical,
\begin{align}
	\frac{Z_{\Theta_i}}{ f_{\Theta_i}(Y)} \widetilde{P}_Y\left( Y; \Theta_i \right) = \frac{Z_{\Theta_j}}{ f_{\Theta_j}(Y)  } \widetilde{P}_Y\left( Y; \Theta_j \right).
   \label{eq:continuity_constraint}
\end{align} 
In case of the unbiased distribution $k=u$ with $\Theta_u=(0,0,0)$ and
$\widetilde{P}_Y( Y ; \Theta_u)=P_Y(Y)$ we know that $Z_{\Theta_u}=1$. As a result
$Z_{\Theta_k}$, $k=1$, $\ldots$, $K$ can be generated according to
Eq.~\eqref{eq:continuity_constraint} in a successive manner from overlapping
distributions starting with $Z_{\Theta_u}=1$. A useful implementation of this process
is described in Ref.~\cite{Schawe2019LargeDeviationsOfConvexHullsETC}. Once the
constants $Z_{\Theta_k}$ have been fixed, the reweighting according to
Eq.~\eqref{eq:bias_correct} is performed and one obtains the final distribution.

\begin{figure}[tb!]
  \begin{center}
    \includegraphics{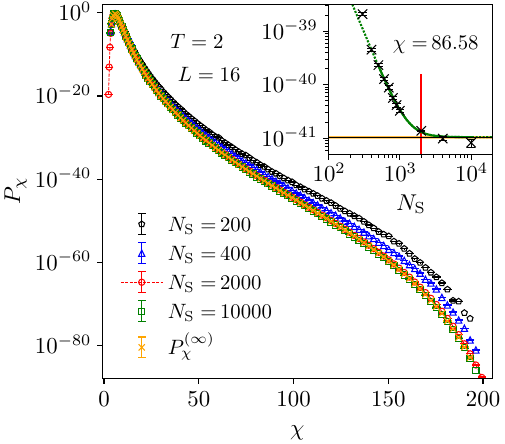}
    \caption{Distribution of the magnetic susceptibility at $T=2$ inside the
      Griffiths phase for $L=16$. The plot illustrates how the measured distribution
      depends on the number $N_\mathrm{S}$ of samples used to compute the estimate of the
      observable $\chi_{\bm{J}}$. The extrapolated curve is obtained by a power-law
      fit $P_\chi(\chi; N_\mathrm{S}) = c_1(\chi)N_\mathrm{S}^{-c_2(\chi)}+ P_{\chi}^{(\infty)}(\chi)$
      at each bin, where $c_1$ and $c_2$ are fit parameters. The inset shows the
      power-law fit for the bin at $\chi=86.58$. The green line corresponds to the
      fit, and its solid part indicates the fit range. The orange shaded area has a
      width of two times the standard error with the value of $P_{\chi}^{(\infty)}$
      at its center. The red vertical line in the inset marks $N_\mathrm{S}=2000$, the number
      of samples that is used throughout the rest of this work.}
    \label{fig:number_of_samples}
  \end{center}
\end{figure}

\begin{figure}
  \begin{center}
    \includegraphics{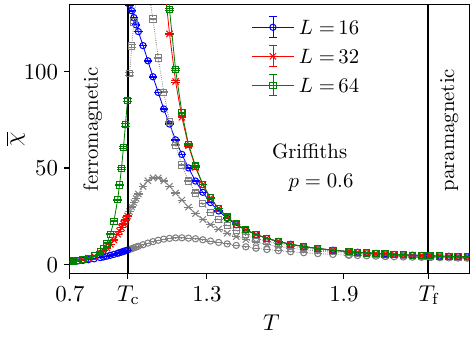}
    \caption{The mean magnetic susceptibility $\overline{\chi}$
      as a function of temperature $T$ for a fraction $p=0.6$ of
      ferromagnetic bonds. The values for the different configurations ${\bm J}$ are
           computed according to the estimator of Eq.~\eqref{eq:cluster_estimator}. The grey lines
           are continuations of the low-temperature cluster estimator into the
           high-temperature phase. These curves exhibit a peak which shifts towards
           the critical temperature on increasing the system size, similar to other
           finite-size definitions of $\overline{\chi}$. The Griffiths phase extends from the
           critical temperature of the pure ferromagnet $T_\mathrm{f}=2.2691 \dots$
           to the corresponding temperature of the diluted system
           $T_\mathrm{c}(0.6)=0.9541(10)$.}
		\label{fig:df_mean_chi_plot}
	\end{center}
\end{figure}

To estimate the value of the magnetic susceptibility for a given bond configuration
it is necessary to compute an average over multiple thermal samples (since the
susceptibility is no ``configurational'' quantity \cite{ebert:22}, i.e., it cannot be estimated from a single spin configuraton). To this end, we
carry out a thermal Monte Carlo simulation utilizing the Swendsen-Wang cluster
algorithm~\cite{SwendsenWang1987NonuniversalCriticalDynamicsInMonteCarloSimulations}. This
method is known to perform well for diluted
ferromagnets~\cite{Kole2022ComparisonOfClusterAlgorithmsForTheBondDilutedIsingModel},
and it is possible to use improved cluster estimators for the
susceptibility~\cite{wolff:89}. Furthermore, the clusters underlying the algorithm
provide an interesting geometrical interpretation of the magnetic susceptibility. To
elucidate this context, in the following we hence provide a short exposition of the
Swendsen-Wang algorithm as well as the corresponding cluster framework. Starting with
a given spin configuration, in the Swendsen-Wang algorithm one \emph{occupies} each ferromagnetic 
bond $J_{\bm{x}\bm{y}}=1$ with probability
$p_{\mathrm{FK}}(J_{\bm{x}\bm{y}})=1-\exp(-2/T)$ if the two connected spins are
parallel, i.e., if $s_{\bm{x}}s_{\bm{y}}=1$, and
$p_{\mathrm{FK}}(J_{\bm{x}\bm{y}})=0$ if $s_{\bm{x}}s_{\bm{y}}= -1$. On the contrary,
diluted bonds with $J_{\bm{x}\bm{y}} = 0$ are never occupied. Two spin sites which
are connected by a path of occupied bonds belong to the same cluster. Clusters which
are defined in this way are denoted as FKCK (Fortuin-Kasteleyn--Coniglio-Klein)
clusters \footnote{The FKCK clusters are also called FK clusters or CK droplets due
  to their historical origin \cite{ConiglioFierro2021CorrelatedPercolation}, see also
  Ref.~\cite{munster:23}.}. The smallest possible FKCK cluster contains only a single
spin site. After constructing the clusters, each of them is randomly assigned an up
or down orientation and the spins in each cluster are flipped accordingly. As a
result all spins within each cluster have identical sign but the sign of two spins in
different clusters may differ. This generates the next spin configuration of the
Monte Carlo process. The Swendsen-Wang algorithm is ergodic and satisfies the
detailed balance condition with respect to the Gibbs-Boltzmann distribution
\cite{NewmanBarkema1999MonteCarloMethodsInStatisticalPhysics}. The cluster estimator
of the magnetic susceptibility is based on the densities $\hat{\rho}_i$ of the
clusters which are defined as their number of sites divided by $N$.  We assume that
the indices $i$ are sorted by cluster size such that $\hat{\rho}_1$ corresponds to
the largest cluster. For $T\geq T_\mathrm{c}$ the estimator is given by the average
cluster size of all FKCK clusters~\cite{wolff:89}. For $T<T_\mathrm{c}$ it is
necessary to subtract the square of the density of the infinite cluster
\cite{BinderHeermann2010SomeImportantRecentDevelopmentsOfTheMCMethology,NewmanBarkema1999MonteCarloMethodsInStatisticalPhysics,
  RoussenqConiglioStauffer1982StudyOfDropletsForCorrelatedSiteBondPercolationIn3D,weigel:02a},
\begin{align}
  \chi_{\bm{J}} =
  \begin{cases}
  \displaystyle
  N \left\langle \sum_{i=1} {\hat{\rho}_i}^{2}  \right\rangle_{\text{FK}}     &\text{if}~T \geq T_\mathrm{c} \\[5mm]
  \displaystyle
  N \left( \left\langle \sum_{i=1}  {\hat{\rho}_i}^{2} \right\rangle_{\text{FK}}  -    \langle \hat{\rho}_1 \rangle_{\text{FK}}^2 \right)&\text{if}~T< T_\mathrm{c}
 \end{cases}.
  \label{eq:cluster_estimator}
\end{align}
The sum is performed over the densities $\hat{\rho}_i$ of all clusters.
In the thermodynamic limit and inside the
ferromagnetic phase there exists a single infinite cluster whose density
$\langle \hat{\rho}_1\rangle_{\text{FK}}$ is equal to the absolute value of the magnetization per site
\cite{ConiglioFierro2021CorrelatedPercolation}. In the numerically studied
finite-size systems we have taken the cluster of largest size as a proxy for the infinite cluster.

To numerically estimate $\chi_{\bm{J}}$, we perform a thermal average over
$N_\mathrm{S}$ cluster configurations. Due to the computational complexity of the
problem successive configurations from the Monte Carlo chain are used, such that
these individual estimates are correlated. The computational complexity emerges since
one has to compute a thermal average each time a new bond sample is proposed. While
we need to compute $\chi_{\bm{J}}$ for each bond update, measurements are only recorded for analysis
after each sweep consisting of $2N$ bond updates according to the bond Monte Carlo
algorithm, see Eq.~\eqref{eq:bond_metropolis}. Figure~\ref{fig:ysampling} shows the
sampling of $\chi_{\bm{J}}$ for systems of size $L=10$ with $p=0.6$ and
$N_\mathrm{S}=2000$ samples at temperature $T=2$. The parameters used to generate the
data are listed in Table \ref{tab:ysampling}. As discussed above, $\theta_1 < 0$
leads to a decrease of $\chi$ and $\theta_1 > 0$ leads to an increase of
$\chi_{\bm{J}}$ compared to the unbiased case $\Theta_u=(0,0,0)$. For
$\Theta_3=(0.9,0,0)$ the Monte Carlo chain oscillates between two equilibrium
states. The gap between these states increases with system size such that sampling in
the intermediate region becomes difficult, and the algorithm is not well suited to
study larger systems with
$\theta_2 = \theta_3 =
0$~\cite{Hartmann2011LargeDeviationPropertiesOfLargestComponentForRandomGraphs}. This
problem can be circumvented by using an appropriate combination of non-zero values
for $\theta_2$ and $\theta_3$; this is illustrated in Fig.~\ref{fig:ysampling}.  The
Figure shows the data of 11 parameter sets for a system of size $L=10$. For larger
systems one has to increase the number of parameter sets to sample the distribution
over a wide range of the support such that computations become more and more
time-consuming. To generate the histogram for the largest studied system size
$L=128$, see Fig.~\ref{fig:df_t0_distr_subplot_and_heat_map}, we used around 300
parameter sets. The algorithm for choosing these parameters is to a certain degree heuristic in that the Gaussians are chosen in number and position in such a way as to fill the relevant range of $\chi$ while maintaining sufficient overlap in the histograms of individual simulations that is required for the stitching together of simulations according to Eq.~\eqref{eq:continuity_constraint}.
      
To ensure that the Monte Carlo chain is in equilibrium we initialize the lattice with
no bonds or all bonds present and wait until both processes oscillate around
approximately the same value, within the fluctuations. Only then the sampling is
started.  After such sampling is completed, the data are combined into a single
histogram by using Eq.~\eqref{eq:bias_correct}.

Figure~\ref{fig:number_of_samples} depicts histograms of the distribution of the
magnetic susceptibility at system size $L=16$ with $p=0.6$ and temperature $T=2$. As
is clearly visible, the distributions depend on the number $N_\mathrm{S}$ of spin
configurations that are used to estimate $\chi_{\bm{J}}$. The histograms are expected
to converge to the asymptotic ones for $N_\mathrm{S}\to \infty$. In some cases the convergence
can be described by a power law as is demonstrated in the inset of
Fig.~\ref{fig:number_of_samples}. Because this does not consistently work well for
all parts of the distributions for all investigated temperatures, we finally settled
on using $N_\mathrm{S}=2000$ as a trade-off between computational effort and accuracy
throughout the rest of this article. The error bars of the distributions are computed
by bootstrapping as described in
Ref.~\cite{Young2015EverythingYouWantedToKnowAboutDataAnalysis}.

\section{Results for the disordered phase and the critical point}
\label{sec:results_griffiths}

We focus our simulations of the two-dimensional bond-diluted Ising ferromagnet on the
case of a fraction of ferromagnetic bonds of $p=0.6$, which is sufficiently far away
both from the pure model as well as from the percolation point. We employ periodic
boundary conditions along both axes. For reference, in
Fig.~\ref{fig:df_mean_chi_plot} we display the temperature dependence of the disorder
average of the magnetic susceptibility,
\begin{align}
    \overline{\chi} = \left[  \chi_{\bm{J}} \right]_J,
\end{align}
estimated 
according to Eq.~\eqref{eq:cluster_estimator}. The critical temperature of the system
is obtained by a finite-size scaling analysis of the wrapping probabilities of the
FKCK clusters, resulting in the estimate $T_\mathrm{c}(0.6)=0.9541(10)$ which is
consistent with previous
works~\cite{ZhongEtAl2020SuperSlowingDownInTheBondDilutedIsingModel,Ohzeki2009LocationsOfMulticriticalPointsForSpinGlassesOnRegularLattices},
for details see Appendix~\ref{sec:critical_temperature_estimate}.  The Griffiths
phase extends from the critical temperature of the pure ferromagnet
$T_\mathrm{f}=2.2691 \dots$ down to $T_\mathrm{c}(0.6)=0.9541(10)$.

\begin{figure}
  \begin{center}
    \includegraphics{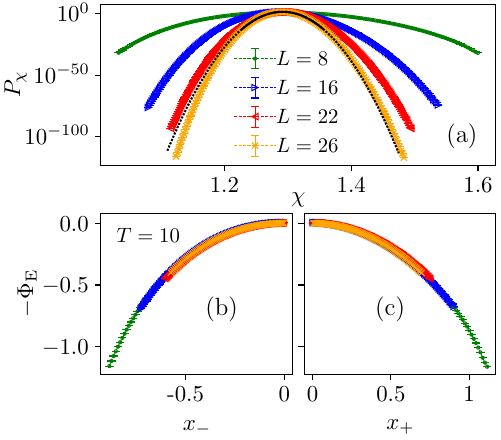}
    \caption{Histogram of the magnetic susceptibility at temperature
      $T=10$, deep inside the paramagnetic phase. (a) Histogram of
      $\chi$ on a logarithmic scale for different system sizes. The
      full black line on top of the orange curve for $L=26$
      corresponds to a Gaussian fit of type $f(x)=c_1 \exp[ -c_2
        (x-c_3)^2 ]$ where $c_1$, $c_2$ and $c_3$ are fit parameters,
      and the dotted black line is an extrapolation of this fit. Close
      to the mean the fit approximates the data quite well. Panels (b)
      and (c) depict the empirical rate function as defined in
      Eq.~\eqref{eq:empirical_rate_function} by using a rescaling of
      the argument according to
      Eq.~\eqref{eq:argument_rescaling_above_tc}. In both cases the
      data collapse onto a single curve. The mean value
      $\overline{\chi}=1.291\,359(22)$ is independent of system size
      within error bars, as is $\chi_\mathrm{f}=1.5670(23)$.}
    \label{fig:df_t10_distr_subplot}
  \end{center}
\end{figure}

For $T > T_\mathrm{f}$ the distribution of the magnetic susceptibility is expected to
be fully concentrated around its mean with a width that decreases by increasing the
system size and that ultimately becomes zero in the thermodynamic limit. In order to
test these predictions we extracted the distribution of the magnetic susceptibility
at $T=10$ deep inside the paramagnetic phase. Figure~\ref{fig:df_t10_distr_subplot}
shows this distribution for a range of various system sizes. The width of the
distribution decreases with system size. The shape of the distribution in the
vicinity of the mean can be approximated by a Gaussian.

Further away from the mean the theory of large deviations provides an appropriate
toolbox to describe a
distribution~\cite{Touchette2009TheLargeDeviationApproachToStatisticalMechanics}. A
central element of large deviations theory is the rate function $\Phi$ that
characterizes the probabilities of exponentially rare events. Following this
approach, the exponential tail of the distribution of an intensive quantity
$x$ which
satisfies the so-called large-deviation principle can be written as
\begin{align}
	P_x(x;N) =  \exp\left\{ -\Phi(x)N +o(N) \right\}
\end{align} 
in the limit $N \to \infty$, where $o(N)$ corresponds to the ``small $o$
notation''. Hence, the fact that the large-deviation principle holds means that the
size dependence of $P_x(x;N)$ on $N$ can be separated from the dependence on $x$.
For finite systems and in case of sufficiently fast convergence the empirical rate
function
\begin{align}
 \Phi_{\mathrm{E}}(x;N)= - \frac{1}{N} \ln\left\{ P_{x}(x;N) \right \}  
    \label{eq:empirical_rate_function}
\end{align}
will provide a good approximation of $\Phi$. The magnetic susceptibility is an
extensive quantity. Therefore it is necessary to perform a rescaling to obtain a well
defined rate function. While a simple rescaling with $N$ or the mean
$\overline{\chi}$ already yields an intensive quantity, we propose the following,
somewhat more subtle, rescaling that takes into account the asymmetry of the
distribution of $\chi$ and uses a different scaling factor to the left and
to the right of the mean,
\begin{align}
  \displaystyle
  x=\begin{cases}
    x_{-} =  \dfrac{\chi -\overline{\chi}}{\overline{\chi}-1}&\text{if}~\chi < \overline{\chi} \\[3mm]
    x_{+} = \dfrac{\chi -\overline{\chi}}{ \chi_\mathrm{f} -\overline{\chi} }&\text{if}~\chi \geq \overline{\chi} \\  
  \end{cases}
  \label{eq:argument_rescaling_above_tc}
\end{align}
which we apply for $T \geq T_\mathrm{c}$. If $\chi < \overline{\chi}$ we divide by
$\overline{\chi}-1$ because $\chi=1$ is the minimum value which the magnetic
susceptibility can assume for $T \geq T_\mathrm{c}$ as one can see from
Eq.~\eqref{eq:cluster_estimator}. When $\chi \geq \overline{\chi}$ we divide by
$\chi_\mathrm{f}-\overline{\chi}$ where $\chi_\mathrm{f}$ corresponds to the 
magnetic susceptibility of the pure ferromagnet in the high-temperature phase,
\begin{align}
\chi_\mathrm{f}:=N \langle \hat{m}^2\rangle_S~~\text{for}~p=1.
\label{eq:m2_ferromagnet}
\end{align}
By doing so we assume that $\chi_\mathrm{f}$ scales in the same way as the largest relevant
values of $\chi$ at the same temperature.

\begin{figure}
  \begin{center}
    \includegraphics{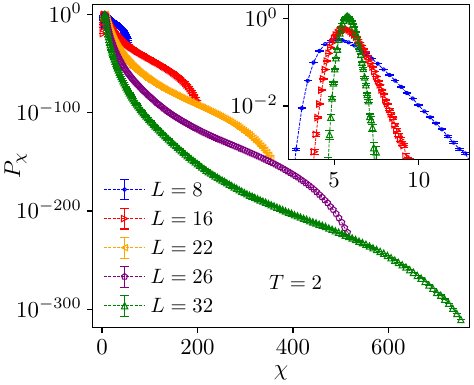}
    \caption{Histogram of the magnetic susceptibility inside the
      Griffiths phase at $T=2$ for multiple system sizes. Within error
      bars and for $L\ge 16$, the mean is independent of system size
      and attains the values $\overline{\chi}=5.8304(11)$. The
      exponential tail is clearly visible. The inset shows histograms
      which are obtained by sampling from the unbiased bond
      distribution, see Eq.~\eqref{eq:bond_ensemble}. For system sizes
      $L=8$ (blue circles) $N_J=10^6$ bond samples are used, for
      systems size $L=16$ (red right-pointing triangles) $N_J=6\times
      10^5$ and for $L=32$ (green upward-pointing triangles) $N_J=1.1
      \times 10^5$. The data demonstrate that using the standard
      sampling approach it is only possible to sample a tiny region
      close to the mean of the distribution.}
    \label{fig:df_t2_distr_plot}
  \end{center}
\end{figure}

\begin{figure}[tb!]
  \begin{center}
    \includegraphics{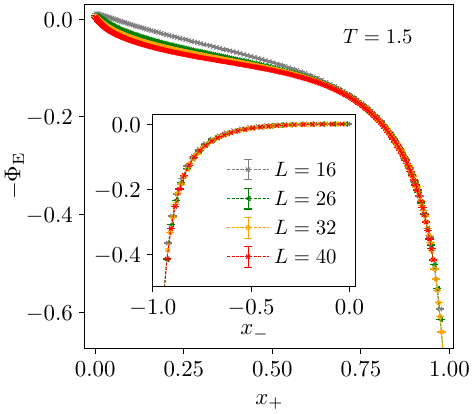}
    \caption{The empirical rate function at $T=1.5$ deep inside the Griffiths phase
      for various system sizes. For $x<0$ the data collapse well onto a single curve
      and for $x>0$ the collapse becomes good for the studied system sizes when $x$
      is larger than approximately $0.75$. The mean is almost independent of system
      size with $\overline{\chi}=16.057(15)$ for $L=16$ and
      $\overline{\chi}=16.173(8)$ for $L=32$ while $\chi_{f}$ diverges as
      $\chi_\mathrm{f} \sim N$ since $T < T_\mathrm{f}$.}
    \label{fig:df_t1x5_ratefunc_plot}
  \end{center}
\end{figure}

\begin{figure}[tb!]
	\begin{center}
         \includegraphics{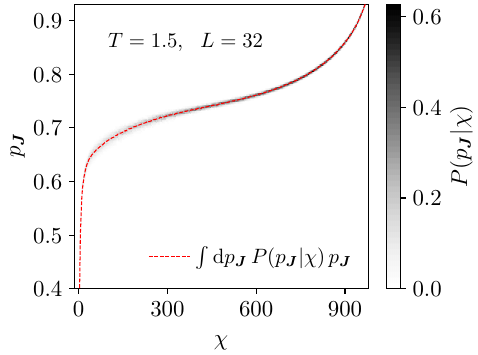}
         \caption{Connection between the value $\chi$ of the magnetic susceptibility
           and the fraction $p_{\bm{J}}$ of ferromagnetic bonds at temperature
           $T=1.5$ and system size $L=32$. The heat map illustrates the probability
           to measure the fraction of ferromagnetic bonds $p_{\bm{J}}$ given the
           magnetic susceptibility $\chi$, denoted as $P(p_{\bm{J}}| \chi)$. The
           dotted red line corresponds to the conditional mean value of the fraction
           of ferromagnetic bonds.}
		\label{fig:df_chi_vs_p_correlation_map_plot}
	\end{center}
\end{figure}

\begin{figure}[tb!]
	\begin{center}
         \includegraphics{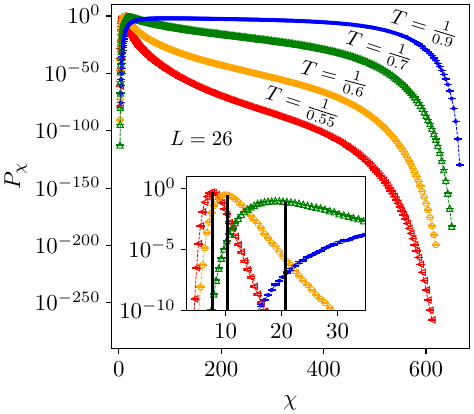}
         \caption{Histogram of the magnetic susceptibility for different temperatures
           $T$ in the Griffiths phase. The temperatures reach from the higher
           temperature Griffiths phase just below $T_\mathrm{f}$ at
           $T=1/0.55=1.\overline{81}$ down to $T=1/0.9=1.\overline{1}$, just above
           the ferromagnetic transition point at $T_\mathrm{c}(0.6)=0.9541(10)$. The
           inset shows the histograms close to the mean. The black vertical lines
           mark the values of the means.}
		\label{fig:df_above_tc_distr_plot}
	\end{center}
\end{figure}

\begin{figure}
	\begin{center}
         \includegraphics{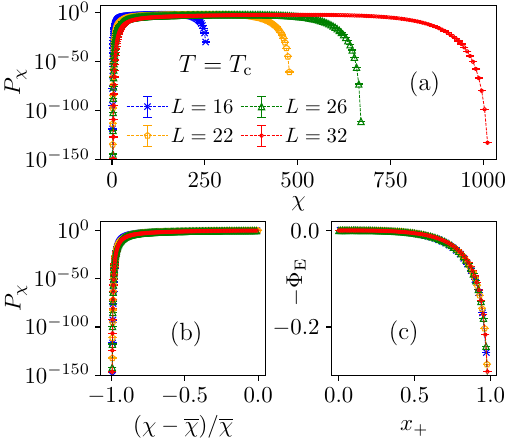}
         \caption{Histogram of the magnetic susceptibility at the critical
           temperature. (a) Distribution of $\chi$ for multiple system sizes on a
           logarithmic scale. (b) Collapse of the data left of the mean by a
           rescaling of the $x$-axis, demonstrating the lack of self-averaging. (c)
           Empirical rate function right of the mean.}
		\label{fig:df_tc_distr_subplot}
	\end{center}
\end{figure}

Figure~\ref{fig:df_t10_distr_subplot} shows the empirical rate function of $\chi$ at
$T=10$. The data collapse quite well onto a single curve, although some corrections
are visible that we expect to diminish further as the system size is
increased. Nevertheless, we presume that the empirical rate function gives a good
impression of the shape of the rate function for $N\to \infty$, indicating that the
assumed large-deviation principle is satisfied.

Next we consider the distribution of the magnetic susceptibility inside the Griffiths
phase. Figure~\ref{fig:df_t2_distr_plot} illustrates the distribution at temperature
$T=2$ for various system sizes. As in the previous case with $T=10$, on increasing
the system size the distribution contracts around the mean, thus demonstrating the
presence of self averaging in the model \cite{NishimoriOrtiz2011RandomSystems}. Also,
the mean $\overline{\chi}$ is essentially independent of system size, but compared to
the distribution at $T=10$ the distribution has acquired a long tail that extends
over a wide range of the support.

As discussed above, the analysis of Bray~\cite{Bray1987NatureOfTheGriffithsPhase}
predicts that as a consequence of the Griffiths singularity the tail of the
distribution of the magnetic susceptibility extends to infinity. To check this
prediction, the rate function is investigated deep inside the Griffiths
phase. Figure~\ref{fig:df_t1x5_ratefunc_plot} shows the empirical rate function at
$T=1.5$. Again, the mean is essentially independent of system size. The rescaling
according to Eq.~\eqref{eq:argument_rescaling_above_tc} leads to a good data collapse
on the left side of the mean. On the other hand, to the right of the mean the data
collapse becomes better with increasing $x_{+}$ such that at $x_{+} \approx 0.75$ all
curves start to fall on top of each other. Note that since $T< T_\mathrm{f}$,
$\chi_\mathrm{f}$ scales as $\chi_\mathrm{f}\sim N$. As a consequence, the empirical
rate function demonstrates that the exponential tail does not become smaller with
system size but will reach into infinity for $N \to \infty$. The data are therefore
fully consistent with the expected behavior resulting from the Griffiths singularity.

Bray~\cite{Bray1987NatureOfTheGriffithsPhase} also derived a lower bound for the
exponential tail, resulting in the functional form of
Eq.~\eqref{eq:exponential_tail_bray}. Hence, the corresponding rate function is given
by $\Phi(x)=A x$, $x>0$, see also
Appendix~\ref{sec:exponential_tail_bray}. Qualitatively, the data are consistent with
an exponential tail, but the exponent does not seem to be a purely linear
function. The rate function is linear if its second derivative is zero. In case of
$L=40$, for instance, the second derivative of the empirical rate function changes
its sign in the range of $0.25\leq x_{+} \leq 0.5$ but it is not zero in the whole
interval, cf.\ Fig.~\ref{fig:df_t1x5_ratefunc_plot}. If there was an interval where
the second derivative was zero and this interval was to increase in size as
$N\to \infty$, this would imply a partially linear rate function in agreement with
the prediction of Bray. Unfortunately, the data for the relatively small system sizes
studied here do not provide a clear indication for such a behavior.

Large values in $\chi$ are expected to be linked to compact structures of
ferromagnetic bonds with a higher fraction of present bonds~\cite{Bray1987NatureOfTheGriffithsPhase}. The fraction $p_{\bm{J}}$ of
ferromagnetic bonds for a given bond sample $\bm{J}$ can be formally
expressed as
\begin{align}
    p_{\bm{J}}= \frac{1}{2N} \sum_{\langle \bm{x} ,\bm{y} \rangle }J_{ \bm{x} \bm{y}}.
    \label{eq:fraction_of_ferromagnetic_bonds}
\end{align}
In Fig.~\ref{fig:df_chi_vs_p_correlation_map_plot} we show the correlation
between the fraction of ferromagnetic bonds $p_{\bm{J}}$ and the magnetic
susceptibility $\chi$ by using the conditional probability
$P(p_{\bm{J}}| \chi )$. As one can see, the fraction of ferromagnetic bonds
and the value of $\chi$ are indeed correlated. Large values of $\chi$
are connected to large values of $p_{\bm{J}}$ and
small values of $\chi$ to small values of $p_{\bm{J}}$, respectively. This
confirms the predictions.

Another relevant aspect concerns the temperature dependence of the probability
distribution of susceptibilities for
$T>T_\mathrm{c}$. Figure~\ref{fig:df_above_tc_distr_plot} illustrates the
distribution for multiple temperatures above $T_\mathrm{c}$. By lowering the
temperature the mean of the distribution shifts to larger values. More weight
relocates into the tail of the distribution and it becomes flatter. This is in
qualitative agreement with the predictions of Bray~\cite{Bray1987NatureOfTheGriffithsPhase}.

\begin{figure}
	\begin{center}
         \includegraphics{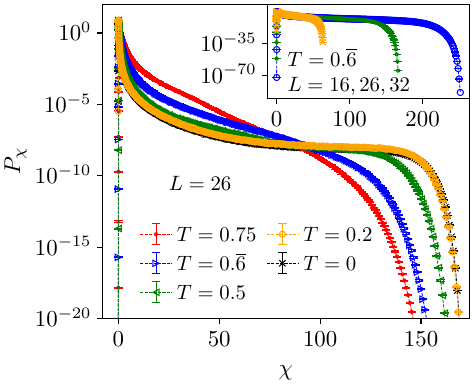}
         \caption{Histogram of the magnetic susceptibility for different temperatures
           below the ferromagnetic phase transition at system size $L=26$. The
           temperatures reach from $T=0.75$, which is relatively close to the phase
           transition, down to the ground state at $T=0$. The inset shows the
           magnetic susceptibility at $T=1/1.5=0.\overline{6}$ for three different
           system sizes $L=16$, $L=26$ and $L=32$. It demonstrates how the tail of
           the distribution widens rapidly with increasing system size.}
		\label{fig:df_below_tc_distr_plot}
	\end{center}
\end{figure}

\begin{figure}
	\begin{center}	\includegraphics{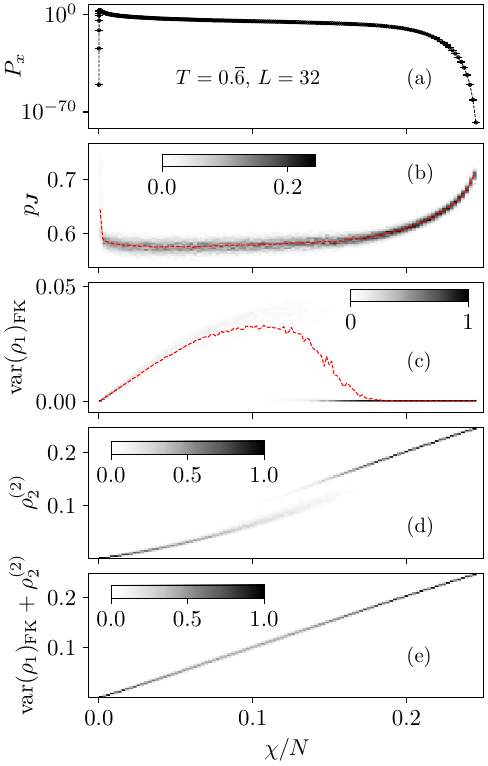}
          \caption{Histogram of the magnetic susceptibility at
            $T=1/1.5=0.\overline{6}$ inside the ferromagnetic phase and heat maps of
            various conditional observables. All plots share the same $x$-axis which
            corresponds to the magnetic susceptibility divided by $N$. (a) Histogram
            of the magnetic susceptibility. (b) Connection between the fraction of
            ferromagnetic bonds and the value of the susceptibility expressed through
            the conditional probability $P(p_{\bm{J}} | \chi )$. The dotted red line
            is the conditional mean and shall serve as a guide to the eye. (c)
            Connection of $\chi$ to the variance of the largest FKCK cluster,
            $P( \mathrm{var}(\rho_1)_{\text{FK}} |\chi )$. (d) Influence of the
            second moment of the second largest FKCK cluster, $\rho_2^{(2)}$, on
            $\chi$, i.e., $P ( \rho_2^{(2)} | \chi )$. (e) Distribution
            $P( \mathrm{var}(\rho_1)_{\text{FK}}+\rho_2^{(2)} | \chi)$, demonstrating
            that
            $\chi \approx N \left(\mathrm{var}(\rho_1)_{\text{FK}}+\rho_2^{(2)}
            \right)$.}
		\label{fig:df_beta1x5_distr_correlation_map_subplot}
	\end{center}
\end{figure}  

Finally, Fig.~\ref{fig:df_tc_distr_subplot} shows the distribution of $\chi$ at the
critical temperature. The distribution has a concave shape, it is very flat and
covers the whole range of the support from $\chi=1$ to $\chi_f\sim N$. The mean of
the distribution diverges according to $\overline{\chi} \propto L^{\gamma_f/\nu_f}[1+\mathcal{O}(1/\ln L)]$,
where $\gamma_f=7/4$ and $\nu_f=1$ are the critical exponents of the pure ferromagnet and $\mathcal{O}(1/\ln L)$ refers to 
the ``big $\mathcal{O}$ notation''~\cite{HasenbuschEtAl2008UniversalDependenceOnDisorderOf2DRandomlyDilutedAndRandomBondPMJIsingModels}. To
the left of the mean there is no convergence to a rate function. Instead a good data
collapse is obtained by only rescaling the $x$-axis according to
$(\chi-\overline{\chi})/\overline{\chi}$.
The origin of this behavior is the lack of
self averaging at criticality in case of the diluted ferromagnet
\cite{WisemanDomany1995LackOfSelfAveragingInCriticalDisorderedSystems,WisemanDomany1998FSSAndLackOfSelfAveragingInCriticalDisorderedSystems,AharonyHarris1996AbsenceOfSelfAveragingAndUniversalFluctuationsInRandomSystemsNearCriticalPoints}. To
the right of the mean the large-deviation principle seems to be satisfied and one
obtains a good collapse of the data by using a scaling rule according to
Eq.~\eqref{eq:argument_rescaling_above_tc}.

\section{Results for the ferromagnetic phase and at zero temperature}
\label{sec:results_below_tc}

\begin{figure}
  \begin{center}
    \includegraphics{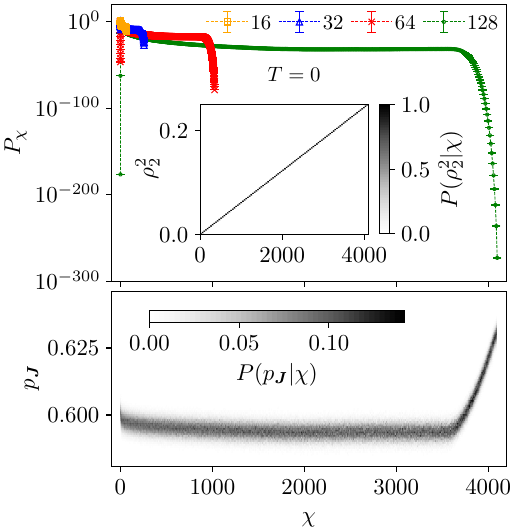}
    \caption{Histogram of the magnetic susceptibility and connected quantities
      at $T=0$. The main plot at the top shows the histogram of the magnetic
      susceptibility for multiple system sizes. The inset is a heat map which
      visualizes the conditional probability to obtain a certain density of the
      second largest cluster $\rho_2$ given the magnetic susceptibility $\chi$,
      $P( \rho_2^2 | \chi )$. The linear relation between $\chi$ and $\rho_2^2$
      is clearly visible. The heat map at the bottom of the figure shares the
      $x$-axis with the main plot and illustrates the conditional probability to
      obtain the fraction of ferromagnetic bonds $p_{\bm{J}}$ given the magnetic
      susceptibility $\chi$, $P(p_{\bm{J}}|\chi)$, for system size $L=128$.}
    \label{fig:df_t0_distr_subplot_and_heat_map}
  \end{center}
\end{figure}

\begin{figure*}
  (a) \hspace{5.5cm} (b) \hspace{5.5cm} (c) \\
  \includegraphics[scale=0.29]{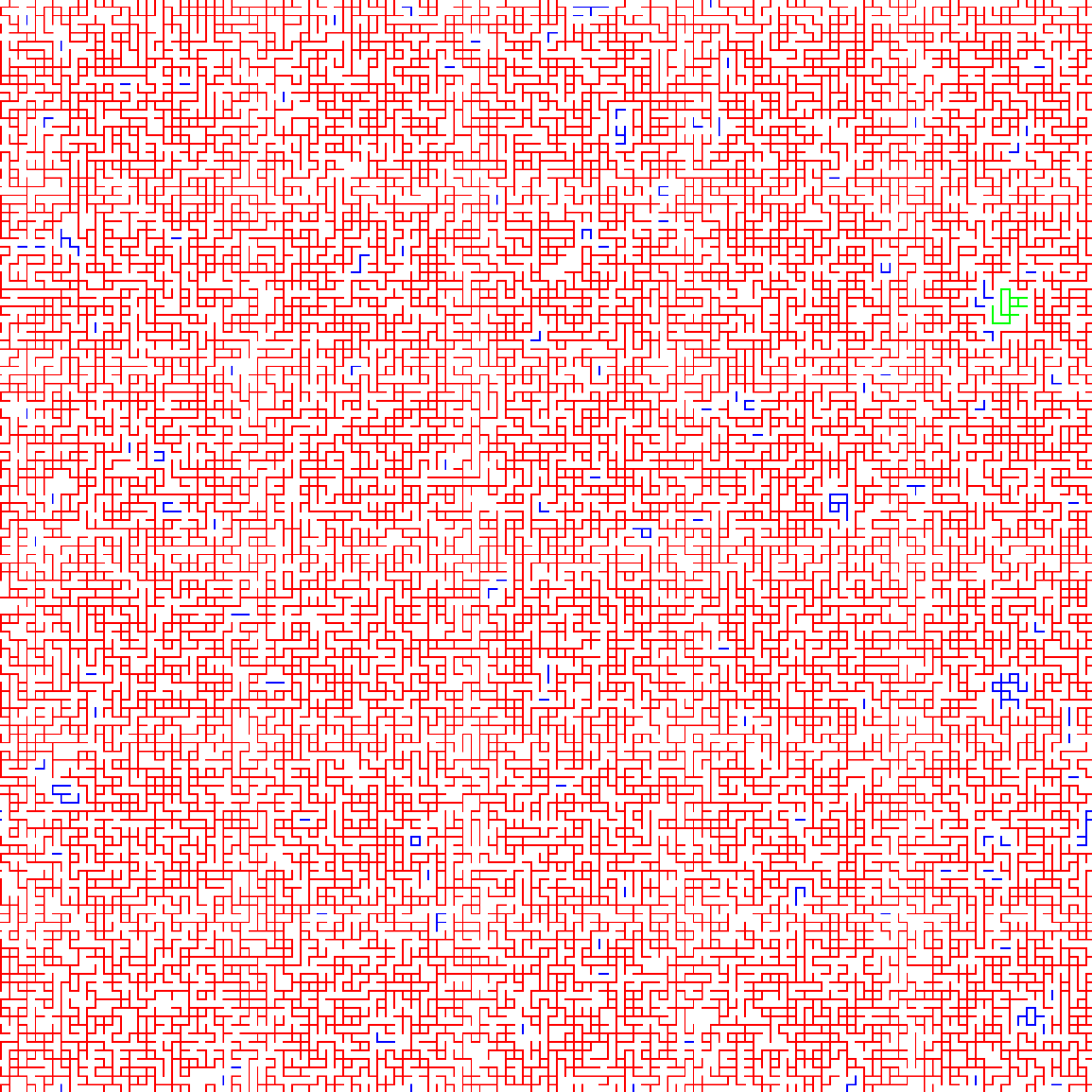}
  \hspace{0.07cm}
  \includegraphics[scale=0.29]{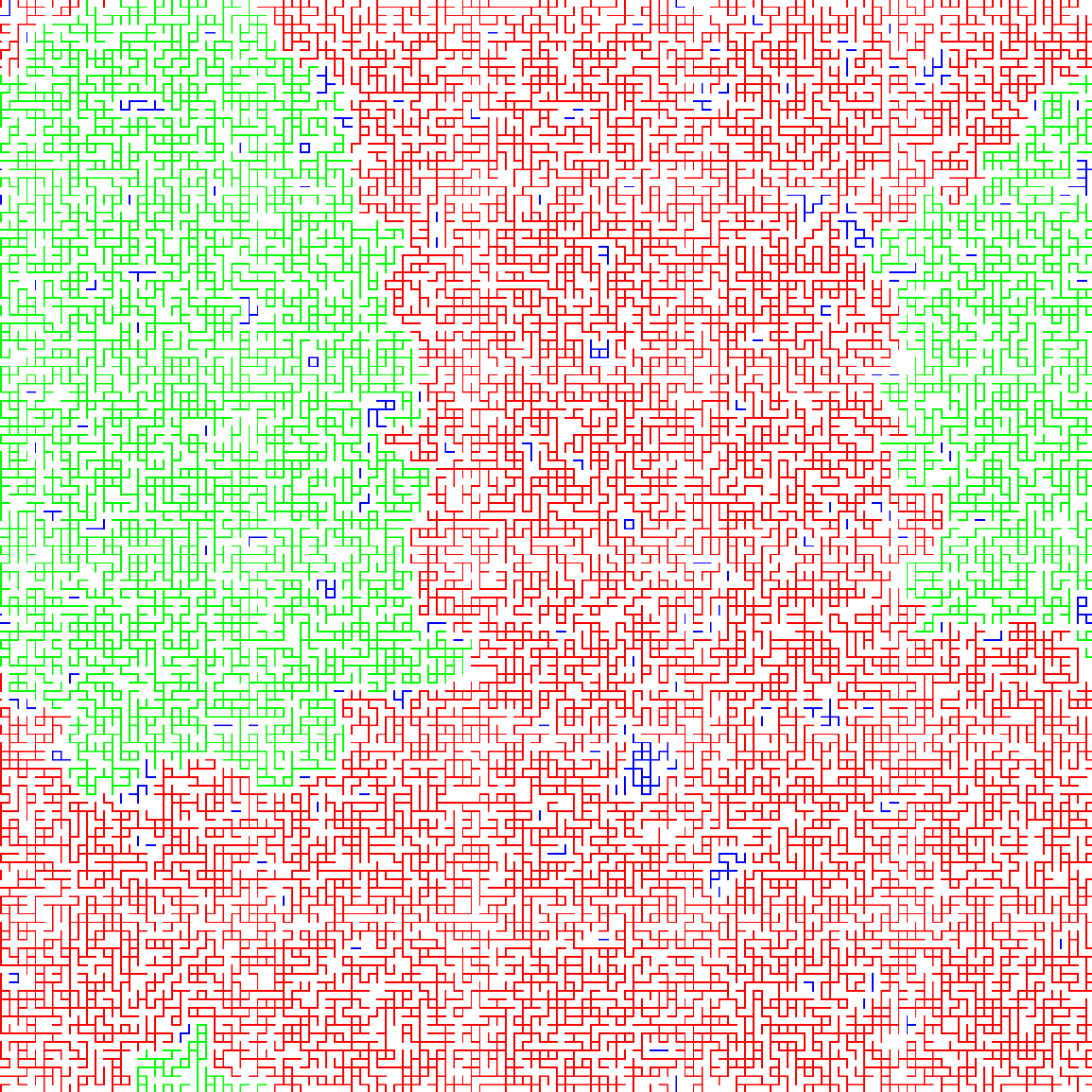}
  \hspace{0.07cm}
  \includegraphics[scale=0.29]{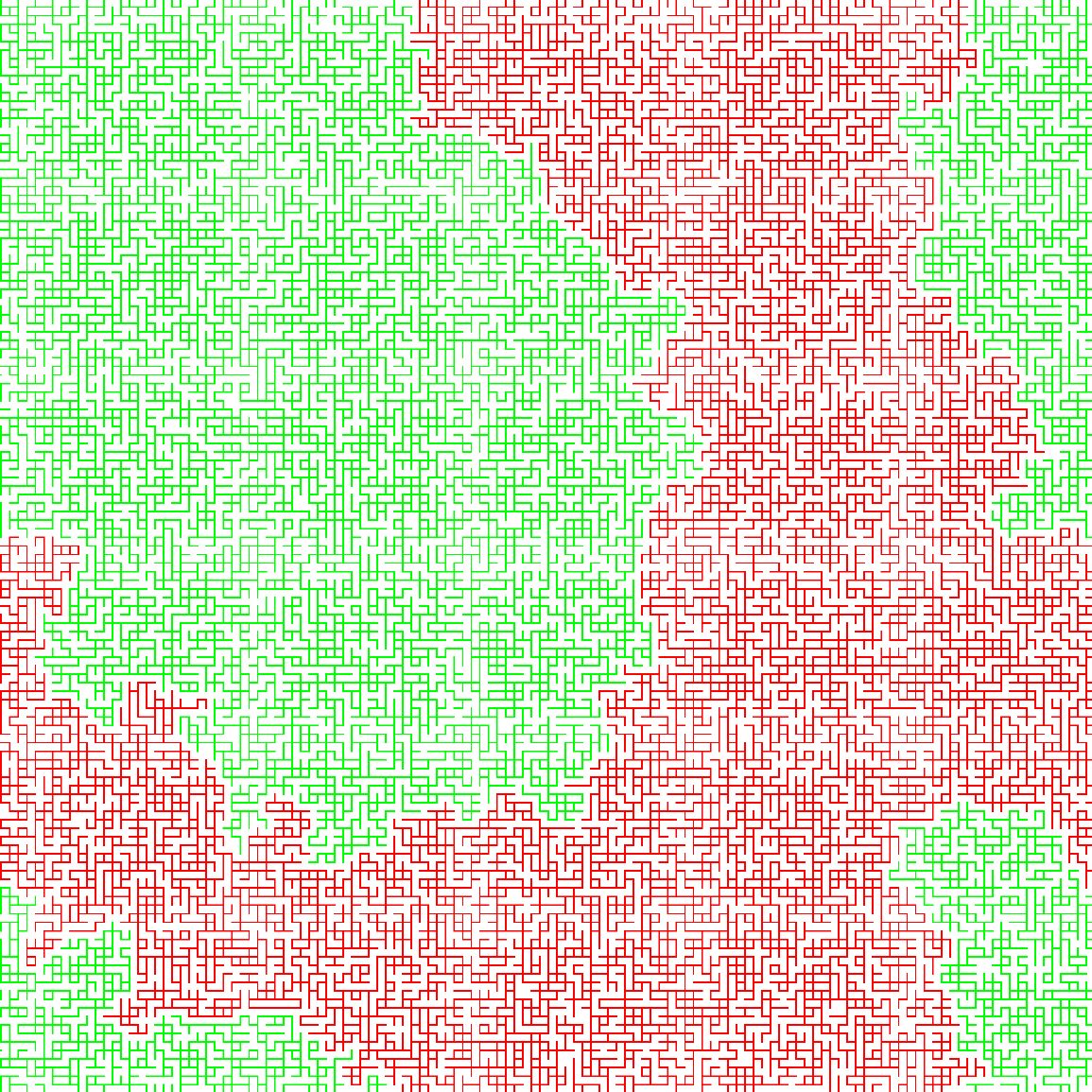}
  \caption{Three bond samples of size $L=128$ with different values of
    $\chi_{\bm{J}}$ at $T=0$. Red bonds belong to the largest cluster, green bonds to
    the second largest cluster, and blue bonds to smaller clusters. Missing bonds are
    white. Bond sample (a) has susceptibility $\chi_{\bm{J}}=0.14233$, which is close
    to the mean value of the distribution $\overline{\chi}= 0.16016(5)$. Bond sample
    (b) originates from the intermediate region of the tail of the distribution with
    $\chi_{\bm{J}}=1519.4$ and bond sample (c) comes from the right end of the tail
    with $\chi_{\bm{J}}=4084.0$. The figures illustrate the significance of the
    second largest cluster for the value of the magnetic susceptibility.}
  \label{fig:df_t0_bond_samples}
\end{figure*}

In this Section we investigate the distribution of the magnetic susceptibility below
the transition temperature $T_\mathrm{c}(p)$ to the ferromagnetic phase~\footnote{As
  a technical detail we want to mention that for the plots in
  Fig.~\ref{fig:df_below_tc_distr_plot},
  Fig.~\ref{fig:df_beta1x5_distr_correlation_map_subplot} and
  Fig.~\ref{fig:df_ferromagnetic_bond_correlation_map_plot} we have used the common
  estimator
  $\chi_{\bm{J}}=N (\langle \hat{m}^2 \rangle_S -\langle \vert \hat{m} \vert
  \rangle_S^2 )$ to compute the magnetic susceptibility since this worked better for
  sampling at larger system sizes. The results, however, are fully consistent with
  those of the cluster estimator of Eq.~\eqref{eq:cluster_estimator} for
  $T<T_{\mathrm{c}}$}. Figure~\ref{fig:df_below_tc_distr_plot} shows the
distribution of the susceptibility for multiple temperatures inside the ferromagnetic
phase. Apart from a strong peak close to the mean value there exits an intermediate
range where the distribution only decays relatively slowly before it again starts to
rapidly fall off. This intermediate range in the tail of the distribution becomes
wider and flatter on lowering the temperature such that there exits a pronounced
plateau region for $T=0$. The distribution for $T=0.2$ is already very similar to
that of the ground state behavior found at zero temperature.

In order to unveil the mechanism leading to the large values of the susceptibility in
the tail of the distribution inside the ferromagnetic phase, we investigated the
correlations of such large susceptibilities to various other
observables. Figure~\ref{fig:df_beta1x5_distr_correlation_map_subplot} shows the
histogram of the susceptibility at $T=0.\overline{6}$ [panel (a)] and the connection
to multiple other quantities for system size $L=32$ [panels (b)-(e)]. It is visible
from panel (b) that the local fraction $p_{\bm{J}}$ of ferromagnetic bonds is
relatively constant over a wider range of the distribution with a value that is
slightly smaller than $p=0.6$. Only for the case where $\chi$ is extremely small or
large can one observe an increase in $p_{\bm{J}}$ [panel (b)]. In the intermediate
tail range, the value of the susceptibility is composed of mainly two contributions,
the variance of the largest FKCK cluster,
$\mathrm{var}(\rho_1)_{\text{FK}}= \langle \hat{\rho}_1^{2} \rangle_{\text{FK}}-\langle
\hat{\rho}_1 \rangle_{\text{FK}}^2$ [panel (c)], and the second moment of the second
largest FKCK cluster, $\rho_2^{(2)}=\langle \hat{\rho}_2^{2} \rangle_{\text{FK}}$
[panel (d)], i.e.,
$\chi \approx N \left(\mathrm{var}(\rho_1)_{\text{FK}}+\rho_2^{(2)} \right)$ [panel
(e)], cf.\ the general form shown in Eq.~\eqref{eq:cluster_estimator}. The
contribution of smaller clusters does not seem to be significant.

The contribution of $\mathrm{var}(\rho_1)_{\text{FK}}$ is only relevant relatively
close to the mean of the distribution. The slight decrease of $p_{\bm{J}}$ in this
region is consistent with large values of $\mathrm{var}(\rho_1)_{\text{FK}}$ since
the fraction of ferromagnetic bonds which leads to a critical temperature of
$T_\mathrm{c}(p^{*})=0.\overline{6}$ is roughly 
$p^{*}\approx0.54$ and thus smaller than $p=0.6$. For the very large values of $\chi$
only $\rho_2^{(2)}$ is important. In the region where there is a jump in
$\rho_2^{(2)}$, a second large cluster of ferromagnetic bonds forms, which is not
connected to the rest of the system, see the details in Appendix
\ref{sec:ferromagnetic_bond_cluster}. As will be shown in the following, this
phenomenon can also be directly studied in the zero-temperature distribution of
$\chi$.

This zero-temperature distribution is of special interest since it is not affected by
thermal fluctuations but, instead, it only depends on the bond disorder. Since there
are no thermal fluctuations, the magnetic susceptibility for a fixed realization can
be computed exactly. Only in this limit are the FKCK clusters identical to the
clusters of ferromagnetic bonds since the FKCK occupation probability is one if there
exists a ferromagnetic bond and zero otherwise. The cluster sizes are static and the
magnetic susceptibility is equal to the average cluster size without the largest
cluster. Because it is not necessary to compute thermal averages the computation time
decreases significantly and hence that larger system sizes can be studied.

Figure~\ref{fig:df_t0_distr_subplot_and_heat_map} shows the distribution of the
magnetic susceptibility at zero temperature for multiple system sizes. Again, it is
visible that the tail of the distribution becomes wider on increasing the system
size. Note that in this case the large-deviation approach
 even allows us to sample over the full support of the
distribution. The support extends from the minimal possible value $\chi=0$, when
there is only one cluster which contains all spins, to the maximum value $\chi=N/4$,
when there are two clusters of equal size which together contain all spins. This is a
direct consequence of Eq.~\eqref{eq:cluster_estimator} if one approaches the infinite
cluster by considering the largest cluster of the finite-size systems under
consideration. Figure~\ref{fig:df_t0_distr_subplot_and_heat_map} contains a heat map
which illustrates the relation between the magnetic susceptibility and the local
fraction of ferromagnetic bonds. It shows that the fraction of ferromagnetic bonds
remains almost constant over a wide range of the magnetic susceptibility. Only for
large values of $\chi$ there is a notable increase in $p_{\bm{J}}$. In comparison to
Figure~\ref{fig:df_beta1x5_distr_correlation_map_subplot}, which shows a similar heat
map at $T=1.5$ inside the Griffiths phase, the interval in which $p_{\bm{J}}$ varies
is much smaller. Instead the density of the second largest cluster $\rho_2$ is the
driving force for the values of the magnetic susceptibility. The inset of
Fig.~\ref{fig:df_t0_distr_subplot_and_heat_map} shows a heat map of the conditional
probability $P(\rho_2^2| \chi)$. It demonstrates that there exists a linear relation
between the square of the density of the second largest cluster and the magnetic
susceptibility. This central importance of the second largest cluster can also be
directly visualized by looking at bond samples with distinct different values of
$\chi$ as they are shown in Fig.~\ref{fig:df_t0_bond_samples}. The figure illustrates
that for bond samples with a magnetic susceptibility which is close to the mean value
the size of the second largest cluster is not important. For bond samples which
originate from the plateau region of the distribution, the second largest cluster
becomes significant, and for the extreme tail events there are only two clusters left
with nearly identical size.

\begin{figure}[tb!]
  \begin{center}
    \includegraphics{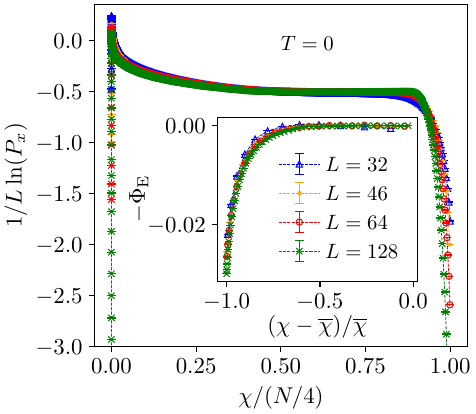}
    \caption{Scaling of the distribution of the magnetic susceptibility at $T=0$ for
      several system sizes. The main plot shows a data collapse of the distribution
      to the right of the mean. Note that the $y$-axis is not scaled according to the
      definition of the empirical rate function in
      Eq.~\eqref{eq:empirical_rate_function}. The chosen scaling gives a good
      collapse within the plateau region of the distribution. The inset shows the
      empirical rate function to the left of the mean,
      $\overline{\chi}=0.16016(5)$. For large system sizes a good collapse of the
      data is obtained.}
    \label{fig:df_t0_rate_function}
  \end{center}
\end{figure}

Finally, the behavior of the zero-temperature distribution with system size is
studied. The main plot in Fig.~\ref{fig:df_t0_rate_function} shows the logarithm of
this distribution divided by $L$. Note that this is a different type of scaling in
comparison to the definition of the empirical rate function given in
Eq.~\eqref{eq:empirical_rate_function}, where one divides by a factor of $N=L^2$. The
scaling with $L$ is a sign of a slower decay behavior of the distribution. The
corresponding data collapse is good in the plateau region of the
distribution. Because the $x$-axis is rescaled by a factor of $N/4$, it is possible
to conclude that the tail of the distribution will extend to infinity in the
thermodynamic limit. The inset of Fig.~\ref{fig:df_t0_rate_function} shows the
empirical rate function to the left of the mean. Note that the zero-temperature
distribution of the magnetic susceptibility is a special feature of the bond-diluted
ferromagnet, i.e., where $J_{\bm{x} \bm{y}}=\{0,1\}$. In the random-bond model with
two types of ferromagnetic bonds of different strengths
\cite{Fisch1978CriticalTemperatureForTwoDimensionalIsingFerromagnetsWithQuenchedBondDisorder},
i.e., where $J_{\bm{x} \bm{y}}=\{c,1\}$ with $0 < c < 1$, the distribution of the
magnetic susceptibility in the thermodynamic limit will be a delta function at the
origin as there is only a single cluster of ferromagnetic bonds which contains all
spin sites \footnote{In case of the random ferromagnet with
  $J_{\bm{x} \bm{y}}=\{c,1\}$ and $0 < c \leq 1$ the variance of the largest FKCK
  cluster might lead to a tail in the distribution of the magnetic susceptibility
  inside the ferromagnetic phase above zero temperature, but whether this is indeed
  the case should be investigated in future studies.}.

Another interesting fact is that the mean of the zero-temperature distribution is
non-zero, $\overline{\chi}=0.16016(5)$. We interpret this as a field-driven continuous phase
transition that emerges at very low temperatures because the finite-size clusters 
will align parallel to the magnetic field, i. e. $\partial \overline{m}/\partial h=\overline{\chi}/T$ such that
$\partial \overline{m}/\partial h \sim T^{-1}$ for $T\to 0$. The external magnetic field destroys the ground state degeneracy.

\section{Discussion}
\label{sec:discussion}

We have studied in depth the distribution of the magnetic susceptibility of the
two-dimensional bond-diluted Ising model for a wide range of different temperature
covering the paramagnetic, the Griffiths, as well as the ferromagnetic phases down to
zero temperature, focusing on a single fraction of ferromagnetic bonds, $p=0.6$. Due
to the adaptation and use of a suitable rare-event sampling algorithm, we are able to
follow the distribution for essentially the full range of the support and down to
probabilities as small as $10^{-300}$. The algorithm is based on the idea proposed in
Ref.~\cite{Hartmann2002SamplingRareEventsStatisticsOfLocalSequenceAlignments} by one
of the present authors to utilize an auxiliary, biased Markov chain in the space of
the disorder degrees of freedom, here the space of coupling configurations. We
combine this approach with the idea of the multiple Gaussian ensemble of
Ref.~\cite{NeuhausHager2006FreeEnergyCalculationsWithMultipleGaussianModifiedEnsembles},
resulting in an efficient algorithm that performs well even for relatively large
systems.

While we cover all phases of the system, our main focus is on the behavior inside the
Griffiths phase, which is the thermal region between the ordering transition in the
pure ferromagnet and the ferromagnetic transition in its diluted counterpart. It is
predicted \cite{Bray1987NatureOfTheGriffithsPhase} that inside the Griffiths phase
the distribution of the magnetic susceptibility has an exponential tail which extends
to infinity, which is a sign of the essential but weak Griffiths singularity. This
singularity is caused by arbitrarily large structures of ferromagnetic bonds in which
the local fraction of ferromagnetic bonds is higher than the average value. Within
these structures the system is effectively in a ferromagnetic state such that a
change of the orientation of the spins can lead to large values of the magnetic
susceptibility. By sampling the distribution of the magnetic susceptibility over a
wide range of the support it is possible to uncover the exponential tail which
emerges inside the Griffiths phase. The connection between the sample fraction of
bonds and the size of the magnetic susceptibility is verified numerically.

The distribution of the magnetic susceptibility is also studied directly at the
critical temperature and inside the ferromagnetic phase.  At the critical temperature
a lack of self-averaging is observed to the left of the distribution mean, i.e., for
small values of $\chi$
\cite{WisemanDomany1995LackOfSelfAveragingInCriticalDisorderedSystems,WisemanDomany1998FSSAndLackOfSelfAveragingInCriticalDisorderedSystems,AharonyHarris1996AbsenceOfSelfAveragingAndUniversalFluctuationsInRandomSystemsNearCriticalPoints}. Inside
the ferromagnetic phase the distribution of the magnetic susceptibility exhibits an
exponential tail similar to that of the Griffiths phase. The tail becomes wider with
increasing system size and it is expected to extend to infinity in the thermodynamic
limit. It is found that the driving force behind large values of the magnetic
susceptibility in the ordered phase is a combination of the variance of the largest FKCK cluster and the second moment of the second largest cluster. At zero-temperature the size of the second largest cluster of ferromagnetic bonds is the only contribution to the large susceptibility observed.

\begin{figure}
  \begin{center}
    \includegraphics{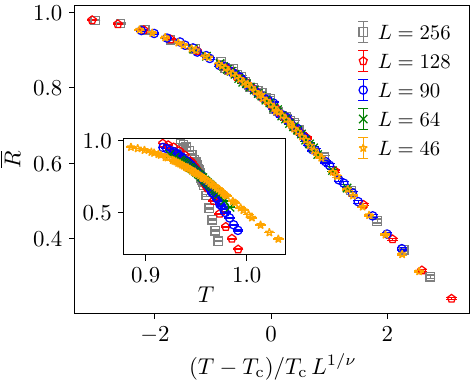}
    \caption{Finite-size scaling of the wrapping probability of FKCK clusters for
      $p=0.6$. The inset shows the original data and the main plot is a data collapse 
      with $T_\mathrm{c}=0.9541(10)$ and $1/\nu=0.90(14)$. To find the
      optimal parameters for the collapse we used the tool provided in
      Ref.~\cite{Melchert2009Autoscale}.
    }	
    \label{fig:df_wrapping_prob_fss_p06}
  \end{center}
\end{figure}  

In the present work, we have focused on the distribution of the susceptibility for a fixed 
expectation value of the fraction of ferromagnetic bonds, namely $p=0.6$. This 
value of $p$ was chosen because it allows for a large Griffiths phase but is still far away from 
the percolation threshold $p_{\text{th}}=0.5$ such that there exist a low temperature ferromagnetic phase.
In general we expect our results to be independent of the value of $p$ in such a sense that an exponential 
tail should exist for all values of $T$ and $p$ inside the Griffiths phase as well as in the ferromagnetic phase, but its shape may look different, see Appendix \ref{sec:results_for_p08} for more details. Furthermore we note that for $T=0$ the magnetic susceptibility
is equal to the average cluster size of the ferromagnetic bond clusters. The problem thereby simplifies to random-bond percolation, suggesting that analytical progress could potentially be made in order to derive the rate function.

The Griffiths phase is not particular to the diluted ferromagnet, but it can also be
observed in other disordered systems, and we hope that the present work will motivate
similar studies of related phenomena in other models. Of particular interest
could be the case of the two-dimensional Ising spin
glass~\cite{MatsudaEtAl2008TheDistributionOfLeeYangZerosAndGriffithsSingularitiesInThePMJModelOfSGs}. This
model exhibits frustrated interactions and does not have a ferromagnetic phase
transition. Thus it would be interesting to see how the distribution of the magnetic
susceptibility is impacted by these differences. Furthermore, we expect that similar
effects of broad distributions will also be visible in observables other than the
magnetic susceptibility. For the diluted ferromagnet studied here, for instance, the
distribution of the specific heat would be of special significance since the specific
heat describes the fluctuations of the internal energy that becomes singular at the
ferromagnetic phase transition just as the magnetic fluctuations represented in the
susceptibility~\cite{MalakisFytas2006LackOfSelfAveragingOfTheSpecificHeatInTheThreeDimensionalRandomFieldIsingModel}. Finally, it would me most intriguing to apply the rare-event sampling
techniques showcased here also for the case of the Griffiths singularities observed
in quantum systems
\cite{Vojta2010QuantumGriffithsEffectsAndSmearedPhaseTransitionsInMetalsTheoryAndExperiment,
  Fisher1992RandomTransverseFieldIsingSpinChains,
  Fisher1995CriticalBehaviorOfRandomTransverseFieldIsingSpinChains,
  YoungRieger1996NumericalStudyOfTheRandomTransverseFieldIsingSpinChain,
  PichEtAl1998CriticalBehaviorAndGriffithsMcCoySingularitiesInThe2DRandomQuantumIsingFerromagnet,
  WangEtAl2017QuantumGriffithsPhaseExperiment,NishimuraEtAl2020GriffithsMcCoySingularityonTheDilutedChimeraGraph}.

\acknowledgments
The authors thank F. Hucht for interesting discussions.  

\appendix

\section{Estimation of the critical temperature}
\label{sec:critical_temperature_estimate}

\begin{figure}[tb!]
  \begin{center}
    \includegraphics{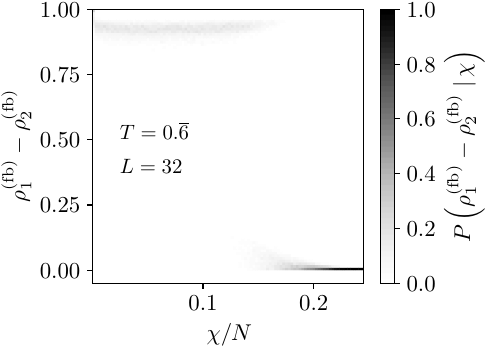}
    \caption{Difference in density of the two largest clusters of ferromagnetic bonds
      for a given value of $\chi$ at temperature $T=0.\overline{6}$ and system size
      $L=32$.}
    \label{fig:df_ferromagnetic_bond_correlation_map_plot}
  \end{center}
\end{figure} 

To determine the critical temperature for $p=0.6$, we performed a finite-size scaling
analysis of the wrapping probabilities $\overline{R}(T)$ of the FKCK clusters. For
our purposes, we define the wrapping probability as the probability that there exists
a connected path of occupied bonds which wraps around the boundaries along the
$x$-axis, the $y$-axis or in both directions. Close to the percolation threshold
finite-size scaling implies that $\overline{R}(T)$ should behave as
\cite{StaufferAharony1994IntroductionToPercolationTheory}
\begin{align}
  \overline{R}(T,L) = f_{\overline{R}} \left \{ \left( T-T_\mathrm{c} \right)/T_\mathrm{c} L^{1/\nu}   \right\},
  \label{eq:crossing-probabilities}
\end{align}
where $f_{\overline{R}}$ is a scaling
function. Figure~\ref{fig:df_wrapping_prob_fss_p06} shows the wrapping probability as
a function of temperature for several system sizes. The number of bond samples used
to perform the average over disorder ranges from $N_J=11\,000$ for the smallest
system size $L=46$ to $N_J=1100$ for the largest system size $L=256$. A collapse of
the data according to Eq.~\eqref{eq:crossing-probabilities} leads to a critical
temperature of $T_\mathrm{c}=0.9541(10)$ and the critical exponent
$1/\nu=0.90(14)$. The reason for the deviation from the value of the pure
ferromagnet, where $\nu_\mathrm{f}=1$, are most likely finite-size scaling
corrections that are not accounted for in the collapse approach. The estimate of the
critical temperature is consistent with previous
results~\cite{ZhongEtAl2020SuperSlowingDownInTheBondDilutedIsingModel,Ohzeki2009LocationsOfMulticriticalPointsForSpinGlassesOnRegularLattices}.


\bigskip

\section{Exponential tail according to Bray}
\label{sec:exponential_tail_bray}

In Ref.~\cite{Bray1987NatureOfTheGriffithsPhase} Bray derived a functional form for
the exponential tail of the distribution of the inverse magnetic susceptibility
$\chi^{-1}$ over the bond disorder inside the Griffiths phase. In the limit
$\chi^{-1} \to 0$ he arrives at the following form for the distribution:
\begin{align*}
    P_{\chi^{-1}}(\chi^{-1})\sim \exp\left(-\frac{A}{\chi^{-1}} \right),
\end{align*}
where $A$ is a temperature dependent positive constant. By performing a change of
variables one obtains the tail as a function of $\chi$,
\begin{align*}
     P_{\chi}(\chi) \sim \chi^{-2} \exp( -A \, \chi ).   
\end{align*}
Below $T_\mathrm{f}$ and outside of the critical region, we have
$\chi_\mathrm{f} \sim N$ such that one can define the intensive quantity $x=\chi/N$
which gives $P_x(x;N) \propto \exp ( -A x N -2 \ln(x)-\ln(N))$.
The corresponding rate
function then is given by
\begin{align}
   \lim_{N \to \infty} -\frac{1}{N} \ln \left\{ P_x(x;N) \right\} = A x.
\end{align}

\section{Ferromagnetic bond clusters}
\label{sec:ferromagnetic_bond_cluster}

Ferromagnetic bond clusters are defined by lattice sites which are connected by a
path of ferromagnetic bonds, i.e., they correspond to the FKCK clusters at $T=0$. The
density of a ferromagnetic bond cluster is given by the number of lattice sites which
it contains divided by $N$.
Figure~\ref{fig:df_ferromagnetic_bond_correlation_map_plot} shows the difference in
density of the two largest clusters of ferromagnetic bonds,
$\rho_1^{(\mathrm{fb})}-\rho_2^{(\mathrm{fb})}$. For $\chi/N \geq 0.2$ the second
largest cluster has almost the same size as the largest cluster, since
$\rho_1^{(\mathrm{fb})}-\rho_2^{(\mathrm{fb})}\approx 0$.  This demonstrates that the
second largest cluster of ferromagnetic bonds leads to the very large values of
$\chi$ inside the ferromagnetic phase at zero and positive temperatures.

\begin{figure}[tb!]
  \begin{center}
    \includegraphics{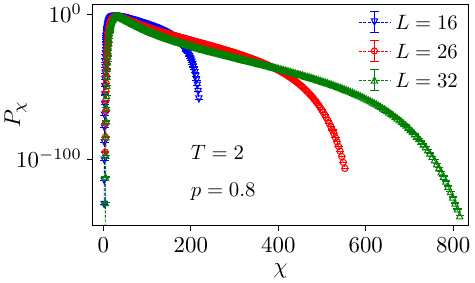}
    \caption{Distribution of the magnetic susceptibility for $p=0.8$ and $T=2$ (inside the Griffiths phase) for different system sizes.}
    \label{fig:df_t2_p08_distr_plot}
  \end{center}
\end{figure} 

\begin{figure}[tb!]
  \begin{center}
    \includegraphics{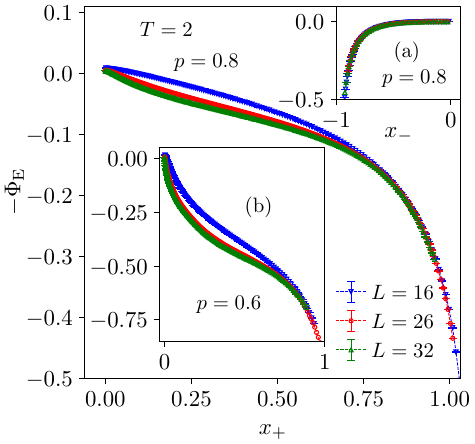}
    \caption{The main plot shows the rate function of the magnetic susceptibility for $p=0.8$ and $T=2$ right of the mean and the inset (a) illustrates the rate function left of the mean. The inset (b) depicts the rate function right of the mean for $p=0.6$ and $T=2$. The $x$-axis is rescaled according to Eq.~\eqref{eq:argument_rescaling_above_tc}.}
    \label{fig:df_t2_p06_p08_ratefunc_plot}
  \end{center}
\end{figure} 

\section{Influence of the occupation probability~$p$}
\label{sec:results_for_p08}
Here we discuss the potential effects of varying the bond occupation probability $p$. The arguments of Bray~\cite{Bray1987NatureOfTheGriffithsPhase}
which explain the existence of an exponential tail in the distribution of magnetic susceptibility inside the Griffiths phase do not depend on a specific value of $p$, but they are 
generally applicable for $0 < p < 1$. The 
shape of the distribution, however, can be different for different values of $p$. Figure~\ref{fig:df_above_tc_distr_plot} demonstrates the temperature dependence of the shape of the distribution.
In a similar way we expect the distribution to look different when the value 
of $p$ is changed, but the presence of an exponential tail inside the Griffiths phase, in general, should not be affected by the specific value of $p$. As an example
Fig.~\ref{fig:df_t2_p08_distr_plot} illustrates the distribution for $p=0.8$ and $T=2$, inside the Griffiths phase. As expected, 
the mean value of $\chi$ is larger than that of $p=0.6$ at the identical temperature, cf.~Fig.~\ref{fig:df_t2_distr_plot}, but the exponential tail is again clearly visible. 
Furthermore, Fig.~\ref{fig:df_t2_p06_p08_ratefunc_plot} depicts the rate function of the distribution and demonstrates that the range of the tail increases with system size. This rate function exhibits the same characteristic shape as the rate function for $p=0.6$ at $T=2$, as shown in the inset (b) for comparison, but it looks more similar to that at temperature $T=1.5$ for $p=0.6$, see Fig.~\ref{fig:df_t1x5_ratefunc_plot}. We have also performed test simulations for $p=0.3$ and again observed an exponential tail in agreement to the theoretical predictions~\cite{Bray1987NatureOfTheGriffithsPhase}. Moreover we have performed a simulation for $T=0$ when $p=0.8$ and found a similar behavior as described in Sec.~\ref{sec:results_below_tc}. We hence find broad numerical confirmation of the expectation that the existence and general featuers of theb Griffiths phase are independent of the bond occupation probability $p$.

\section*{}
\bibliography{griff}

\begin{thebibliography}{69}%
\makeatletter
\providecommand \@ifxundefined [1]{%
 \@ifx{#1\undefined}
}%
\providecommand \@ifnum [1]{%
 \ifnum #1\expandafter \@firstoftwo
 \else \expandafter \@secondoftwo
 \fi
}%
\providecommand \@ifx [1]{%
 \ifx #1\expandafter \@firstoftwo
 \else \expandafter \@secondoftwo
 \fi
}%
\providecommand \natexlab [1]{#1}%
\providecommand \enquote  [1]{``#1''}%
\providecommand \bibnamefont  [1]{#1}%
\providecommand \bibfnamefont [1]{#1}%
\providecommand \citenamefont [1]{#1}%
\providecommand \href@noop [0]{\@secondoftwo}%
\providecommand \href [0]{\begingroup \@sanitize@url \@href}%
\providecommand \@href[1]{\@@startlink{#1}\@@href}%
\providecommand \@@href[1]{\endgroup#1\@@endlink}%
\providecommand \@sanitize@url [0]{\catcode `\\12\catcode `\$12\catcode
  `\&12\catcode `\#12\catcode `\^12\catcode `\_12\catcode `\%12\relax}%
\providecommand \@@startlink[1]{}%
\providecommand \@@endlink[0]{}%
\providecommand \url  [0]{\begingroup\@sanitize@url \@url }%
\providecommand \@url [1]{\endgroup\@href {#1}{\urlprefix }}%
\providecommand \urlprefix  [0]{URL }%
\providecommand \Eprint [0]{\href }%
\providecommand \doibase [0]{https://doi.org/}%
\providecommand \selectlanguage [0]{\@gobble}%
\providecommand \bibinfo  [0]{\@secondoftwo}%
\providecommand \bibfield  [0]{\@secondoftwo}%
\providecommand \translation [1]{[#1]}%
\providecommand \BibitemOpen [0]{}%
\providecommand \bibitemStop [0]{}%
\providecommand \bibitemNoStop [0]{.\EOS\space}%
\providecommand \EOS [0]{\spacefactor3000\relax}%
\providecommand \BibitemShut  [1]{\csname bibitem#1\endcsname}%
\let\auto@bib@innerbib\@empty
\bibitem [{\citenamefont {Brush}(1967)}]{Brush1967HistoryOfTheLenzIsingModel}%
  \BibitemOpen
  \bibfield  {author} {\bibinfo {author} {\bibfnamefont {S.~G.}\ \bibnamefont
  {Brush}},\ }\bibfield  {title} {\bibinfo {title} {History of the
  {L}enz-{I}sing model},\ }\href@noop {} {\bibfield  {journal} {\bibinfo
  {journal} {Rev. Mod. Phys.}\ }\textbf {\bibinfo {volume} {{39}}},\ \bibinfo
  {pages} {883} (\bibinfo {year} {1967})}\BibitemShut {NoStop}%
\bibitem [{\citenamefont {Coniglio}\ and\ \citenamefont
  {Fierro}(2021)}]{ConiglioFierro2021CorrelatedPercolation}%
  \BibitemOpen
  \bibfield  {author} {\bibinfo {author} {\bibfnamefont {A.}~\bibnamefont
  {Coniglio}}\ and\ \bibinfo {author} {\bibfnamefont {A.}~\bibnamefont
  {Fierro}},\ }\bibinfo {title} {Correlated percolation},\ in\ \href@noop {}
  {\emph {\bibinfo {booktitle} {{Complex Media and Percolation Theory}}}},\
  \bibinfo {editor} {edited by\ \bibinfo {editor} {\bibfnamefont
  {M.}~\bibnamefont {Sahimi}}\ and\ \bibinfo {editor} {\bibfnamefont {A.~G.}\
  \bibnamefont {Hunt}}}\ (\bibinfo  {publisher} {Springer},\ \bibinfo {address}
  {New York},\ \bibinfo {year} {2021})\ p.~\bibinfo {pages} {61}\BibitemShut
  {NoStop}%
\bibitem [{\citenamefont {Page}\ and\ \citenamefont
  {Sear}(2006)}]{AmandaSear2006HeterogeneousNucleationInAndOutOfPores}%
  \BibitemOpen
  \bibfield  {author} {\bibinfo {author} {\bibfnamefont {A.~J.}\ \bibnamefont
  {Page}}\ and\ \bibinfo {author} {\bibfnamefont {R.~P.}\ \bibnamefont
  {Sear}},\ }\bibfield  {title} {\bibinfo {title} {Heterogeneous nucleation in
  and out of pores},\ }\href@noop {} {\bibfield  {journal} {\bibinfo  {journal}
  {Phys. Rev. Lett.}\ }\textbf {\bibinfo {volume} {{97}}},\ \bibinfo {pages}
  {065701} (\bibinfo {year} {2006})}\BibitemShut {NoStop}%
\bibitem [{\citenamefont {Grabowski}\ and\ \citenamefont
  {Kosiński}(2006)}]{Grabowski2006IsingBasedModelOfOpinionFormationInAComplexNetworkOfInterpersonalInteractions}%
  \BibitemOpen
  \bibfield  {author} {\bibinfo {author} {\bibfnamefont {A.}~\bibnamefont
  {Grabowski}}\ and\ \bibinfo {author} {\bibfnamefont {R.}~\bibnamefont
  {Kosiński}},\ }\bibfield  {title} {\bibinfo {title} {{I}sing-based model of
  opinion formation in a complex network of interpersonal interactions},\
  }\href@noop {} {\bibfield  {journal} {\bibinfo  {journal} {Physica A}\
  }\textbf {\bibinfo {volume} {{361}}},\ \bibinfo {pages} {651} (\bibinfo
  {year} {2006})}\BibitemShut {NoStop}%
\bibitem [{\citenamefont {Sethna}\ \emph {et~al.}(2001)\citenamefont {Sethna},
  \citenamefont {Dahmen},\ and\ \citenamefont
  {Myers}}]{Sethna2001CracklingNoise}%
  \BibitemOpen
  \bibfield  {author} {\bibinfo {author} {\bibfnamefont {J.~P.}\ \bibnamefont
  {Sethna}}, \bibinfo {author} {\bibfnamefont {K.~A.}\ \bibnamefont {Dahmen}},\
  and\ \bibinfo {author} {\bibfnamefont {C.~R.}\ \bibnamefont {Myers}},\
  }\bibfield  {title} {\bibinfo {title} {Crackling noise},\ }\href@noop {}
  {\bibfield  {journal} {\bibinfo  {journal} {Nature}\ }\textbf {\bibinfo
  {volume} {{410}}},\ \bibinfo {pages} {242} (\bibinfo {year}
  {2001})}\BibitemShut {NoStop}%
\bibitem [{\citenamefont {Stein}\ and\ \citenamefont
  {Newman}(2013)}]{NewmanStein2013SpinGlassesAndComplexity}%
  \BibitemOpen
  \bibfield  {author} {\bibinfo {author} {\bibfnamefont {D.~L.}\ \bibnamefont
  {Stein}}\ and\ \bibinfo {author} {\bibfnamefont {C.~M.}\ \bibnamefont
  {Newman}},\ }\href@noop {} {\emph {\bibinfo {title} {{Spin Glasses and
  Complexity}}}}\ (\bibinfo  {publisher} {Princeton University Press},\
  \bibinfo {address} {Princeton},\ \bibinfo {year} {2013})\BibitemShut
  {NoStop}%
\bibitem [{\citenamefont
  {Nishimori}(2001)}]{Nishimori2001StatisticalPhysicsOfSpinGlassesAndInformationProcessing}%
  \BibitemOpen
  \bibfield  {author} {\bibinfo {author} {\bibfnamefont {H.}~\bibnamefont
  {Nishimori}},\ }\href@noop {} {\emph {\bibinfo {title} {Statistical Physics
  of Spin Glasses and Information Processing: An Introduction}}}\ (\bibinfo
  {publisher} {Oxford University Press},\ \bibinfo {address} {Oxford},\
  \bibinfo {year} {2001})\BibitemShut {NoStop}%
\bibitem [{\citenamefont {Nishimori}\ and\ \citenamefont
  {Ortiz}(2011)}]{NishimoriOrtiz2011RandomSystems}%
  \BibitemOpen
  \bibfield  {author} {\bibinfo {author} {\bibfnamefont {H.}~\bibnamefont
  {Nishimori}}\ and\ \bibinfo {author} {\bibfnamefont {G.}~\bibnamefont
  {Ortiz}},\ }\bibinfo {title} {Random systems},\ in\ \href@noop {} {\emph
  {\bibinfo {booktitle} {Elements of Phase Transitions and Critical
  Phenomena}}}\ (\bibinfo  {publisher} {Oxford University Press},\ \bibinfo
  {address} {Oxford},\ \bibinfo {year} {2011})\ p.\ \bibinfo {pages}
  {178}\BibitemShut {NoStop}%
\bibitem [{\citenamefont {Dotsenko}\ and\ \citenamefont
  {Dotsenko}(1983)}]{Dotsenko1983CriticalBehaviourOfThePhaseTransitionInThe2DIsingModelWithImpurities}%
  \BibitemOpen
  \bibfield  {author} {\bibinfo {author} {\bibfnamefont {V.~S.}\ \bibnamefont
  {Dotsenko}}\ and\ \bibinfo {author} {\bibfnamefont {V.~S.}\ \bibnamefont
  {Dotsenko}},\ }\bibfield  {title} {\bibinfo {title} {Critical behaviour of
  the phase transition in the 2d {I}sing model with impurities},\ }\href@noop
  {} {\bibfield  {journal} {\bibinfo  {journal} {Adv. Phys.}\ }\textbf
  {\bibinfo {volume} {{32}}},\ \bibinfo {pages} {129} (\bibinfo {year}
  {1983})}\BibitemShut {NoStop}%
\bibitem [{\citenamefont
  {Shalaev}(1994)}]{Shalaev1994CriticalBehaviorOfThe2DIsingModelWithRandomBonds}%
  \BibitemOpen
  \bibfield  {author} {\bibinfo {author} {\bibfnamefont {B.}~\bibnamefont
  {Shalaev}},\ }\bibfield  {title} {\bibinfo {title} {Critical behavior of the
  two-dimensional {I}sing model with random bonds},\ }\href@noop {} {\bibfield
  {journal} {\bibinfo  {journal} {Phys. Rep.}\ }\textbf {\bibinfo {volume}
  {{237}}},\ \bibinfo {pages} {129} (\bibinfo {year} {1994})}\BibitemShut
  {NoStop}%
\bibitem [{\citenamefont
  {Harris}(1974)}]{ABHarris1974EffectOfRandomDefectsOnTheCriticalBehaviourOfIsingModels}%
  \BibitemOpen
  \bibfield  {author} {\bibinfo {author} {\bibfnamefont {A.~B.}\ \bibnamefont
  {Harris}},\ }\bibfield  {title} {\bibinfo {title} {Effect of random defects
  on the critical behaviour of {I}sing models},\ }\href@noop {} {\bibfield
  {journal} {\bibinfo  {journal} {J. Phys. C}\ }\textbf {\bibinfo {volume}
  {{7}}},\ \bibinfo {pages} {1671} (\bibinfo {year} {1974})}\BibitemShut
  {NoStop}%
\bibitem [{\citenamefont {Ludwig}\ and\ \citenamefont
  {Cardy}(1987)}]{LudwigCardy1987PerturbativeEvaluationOfTheConformalAnomalyAtNewCriticalPointsWithApplicationsToRandomSystems}%
  \BibitemOpen
  \bibfield  {author} {\bibinfo {author} {\bibfnamefont {A.~W.}\ \bibnamefont
  {Ludwig}}\ and\ \bibinfo {author} {\bibfnamefont {J.~L.}\ \bibnamefont
  {Cardy}},\ }\bibfield  {title} {\bibinfo {title} {Perturbative evaluation of
  the conformal anomaly at new critical points with applications to random
  systems},\ }\href@noop {} {\bibfield  {journal} {\bibinfo  {journal} {Nucl.
  Phys. B}\ }\textbf {\bibinfo {volume} {{285}}},\ \bibinfo {pages} {687}
  (\bibinfo {year} {1987})}\BibitemShut {NoStop}%
\bibitem [{\citenamefont
  {Ludwig}(1987)}]{Ludwig1987CriticalBehaviorOfThe2DRandomQStatePottsModelByExpansionInQMinus2}%
  \BibitemOpen
  \bibfield  {author} {\bibinfo {author} {\bibfnamefont {A.~W.}\ \bibnamefont
  {Ludwig}},\ }\bibfield  {title} {\bibinfo {title} {Critical behavior of the
  two-dimensional random $q$-state {P}otts model by expansion in ($q-2$)},\
  }\href@noop {} {\bibfield  {journal} {\bibinfo  {journal} {Nucl. Phys. B}\
  }\textbf {\bibinfo {volume} {{285}}},\ \bibinfo {pages} {97} (\bibinfo {year}
  {1987})}\BibitemShut {NoStop}%
\bibitem [{\citenamefont
  {Shankar}(1987)}]{Shankar1987ExactCriticalBehaviorOfArandomBondTwoDimensionalIsingModel}%
  \BibitemOpen
  \bibfield  {author} {\bibinfo {author} {\bibfnamefont {R.}~\bibnamefont
  {Shankar}},\ }\bibfield  {title} {\bibinfo {title} {Exact critical behavior
  of a random bond two-dimensional {I}sing model},\ }\href@noop {} {\bibfield
  {journal} {\bibinfo  {journal} {Phys. Rev. Lett.}\ }\textbf {\bibinfo
  {volume} {{58}}},\ \bibinfo {pages} {2466} (\bibinfo {year}
  {1987})}\BibitemShut {NoStop}%
\bibitem [{\citenamefont {Ludwig}(1988)}]{Ludwig1988CommentOnShankar1987}%
  \BibitemOpen
  \bibfield  {author} {\bibinfo {author} {\bibfnamefont {A.~W.~W.}\
  \bibnamefont {Ludwig}},\ }\bibfield  {title} {\bibinfo {title} {Comment on
  "{E}xact critical behavior of a random-bond two-dimensional {I}sing model"},\
  }\href@noop {} {\bibfield  {journal} {\bibinfo  {journal} {Phys. Rev. Lett.}\
  }\textbf {\bibinfo {volume} {{61}}},\ \bibinfo {pages} {2388} (\bibinfo
  {year} {1988})}\BibitemShut {NoStop}%
\bibitem [{\citenamefont {Hasenbusch}\ \emph {et~al.}(2008)\citenamefont
  {Hasenbusch}, \citenamefont {Parisen~Toldin}, \citenamefont {Pelissetto},\
  and\ \citenamefont
  {Vicari}}]{HasenbuschEtAl2008UniversalDependenceOnDisorderOf2DRandomlyDilutedAndRandomBondPMJIsingModels}%
  \BibitemOpen
  \bibfield  {author} {\bibinfo {author} {\bibfnamefont {M.}~\bibnamefont
  {Hasenbusch}}, \bibinfo {author} {\bibfnamefont {F.}~\bibnamefont
  {Parisen~Toldin}}, \bibinfo {author} {\bibfnamefont {A.}~\bibnamefont
  {Pelissetto}},\ and\ \bibinfo {author} {\bibfnamefont {E.}~\bibnamefont
  {Vicari}},\ }\bibfield  {title} {\bibinfo {title} {Universal dependence on
  disorder of two-dimensional randomly diluted and random-bond {$\pm J$}
  {I}sing models},\ }\href@noop {} {\bibfield  {journal} {\bibinfo  {journal}
  {Phys. Rev. E}\ }\textbf {\bibinfo {volume} {{78}}},\ \bibinfo {pages}
  {011110} (\bibinfo {year} {2008})}\BibitemShut {NoStop}%
\bibitem [{\citenamefont {Martins}\ and\ \citenamefont
  {Plascak}(2007)}]{MartinsPlascak2007UniversalityClassOfThe2DSiteDilutedIsingModel}%
  \BibitemOpen
  \bibfield  {author} {\bibinfo {author} {\bibfnamefont {P.~H.~L.}\
  \bibnamefont {Martins}}\ and\ \bibinfo {author} {\bibfnamefont {J.~A.}\
  \bibnamefont {Plascak}},\ }\bibfield  {title} {\bibinfo {title} {Universality
  class of the two-dimensional site-diluted {I}sing model},\ }\href@noop {}
  {\bibfield  {journal} {\bibinfo  {journal} {Phys. Rev. E}\ }\textbf {\bibinfo
  {volume} {{76}}},\ \bibinfo {pages} {012102} (\bibinfo {year}
  {2007})}\BibitemShut {NoStop}%
\bibitem [{\citenamefont {Wang}\ \emph {et~al.}(1990)\citenamefont {Wang},
  \citenamefont {Selke}, \citenamefont {Dotsenko},\ and\ \citenamefont
  {Andreichenko}}]{WangEtAl1990TheCriticalBehaviourOfThe2DDiluteIsingMagnet}%
  \BibitemOpen
  \bibfield  {author} {\bibinfo {author} {\bibfnamefont {J.-S.}\ \bibnamefont
  {Wang}}, \bibinfo {author} {\bibfnamefont {W.}~\bibnamefont {Selke}},
  \bibinfo {author} {\bibfnamefont {V.}~\bibnamefont {Dotsenko}},\ and\
  \bibinfo {author} {\bibfnamefont {V.}~\bibnamefont {Andreichenko}},\
  }\bibfield  {title} {\bibinfo {title} {The critical behaviour of the
  two-dimensional dilute {I}sing magnet},\ }\href@noop {} {\bibfield  {journal}
  {\bibinfo  {journal} {Physica A}\ }\textbf {\bibinfo {volume} {{164}}},\
  \bibinfo {pages} {221} (\bibinfo {year} {1990})}\BibitemShut {NoStop}%
\bibitem [{\citenamefont {Andreichenko}\ \emph {et~al.}(1990)\citenamefont
  {Andreichenko}, \citenamefont {Dotsenko}, \citenamefont {Selke},\ and\
  \citenamefont
  {Wang}}]{AndreichenkoEtAl1990MonteCarloStudyOfThe2DIsingModelWithImpurities}%
  \BibitemOpen
  \bibfield  {author} {\bibinfo {author} {\bibfnamefont {V.}~\bibnamefont
  {Andreichenko}}, \bibinfo {author} {\bibfnamefont {V.}~\bibnamefont
  {Dotsenko}}, \bibinfo {author} {\bibfnamefont {W.}~\bibnamefont {Selke}},\
  and\ \bibinfo {author} {\bibfnamefont {J.-S.}\ \bibnamefont {Wang}},\
  }\bibfield  {title} {\bibinfo {title} {{M}onte {C}arlo study of the 2d
  {I}sing model with impurities},\ }\href@noop {} {\bibfield  {journal}
  {\bibinfo  {journal} {Nucl. Phys. B}\ }\textbf {\bibinfo {volume} {{344}}},\
  \bibinfo {pages} {531} (\bibinfo {year} {1990})}\BibitemShut {NoStop}%
\bibitem [{\citenamefont
  {Hadjiagapiou}(2011)}]{Hadjiagapiou2011MonteCarloAnalysisOfTheCriticalPropertiesOfThe2DRBDIMviaWangLandauAlgorithm}%
  \BibitemOpen
  \bibfield  {author} {\bibinfo {author} {\bibfnamefont {I.~A.}\ \bibnamefont
  {Hadjiagapiou}},\ }\bibfield  {title} {\bibinfo {title} {{M}onte {C}arlo
  analysis of the critical properties of the two-dimensional randomly
  bond-diluted {I}sing model via {W}ang--{L}andau algorithm},\ }\href@noop {}
  {\bibfield  {journal} {\bibinfo  {journal} {Physica A}\ }\textbf {\bibinfo
  {volume} {390}},\ \bibinfo {pages} {1279} (\bibinfo {year}
  {2011})}\BibitemShut {NoStop}%
\bibitem [{\citenamefont {Ikeda}\ \emph {et~al.}(1979)\citenamefont {Ikeda},
  \citenamefont {Suzuki},\ and\ \citenamefont
  {T.~Hutchings}}]{IdekaEtAl1979NeutronScatteringInvestigationOfStaticCriticalPhenomenaInTheTwoDimensionalAntiferromagnets}%
  \BibitemOpen
  \bibfield  {author} {\bibinfo {author} {\bibfnamefont {H.}~\bibnamefont
  {Ikeda}}, \bibinfo {author} {\bibfnamefont {M.}~\bibnamefont {Suzuki}},\ and\
  \bibinfo {author} {\bibfnamefont {M.}~\bibnamefont {T.~Hutchings}},\
  }\bibfield  {title} {\bibinfo {title} {Neutron scattering investigation of
  static critical phenomena in the two-dimensional antiferromagnets:
  {$\mathrm{Rb}_{2} \mathrm{Co}_{c} \mathrm{Mg}_{1-c} \mathrm{F}_{4}$ }},\
  }\href@noop {} {\bibfield  {journal} {\bibinfo  {journal} {J. Phys. Soc.
  Jpn.}\ }\textbf {\bibinfo {volume} {{46}}},\ \bibinfo {pages} {1153}
  (\bibinfo {year} {1979})}\BibitemShut {NoStop}%
\bibitem [{\citenamefont {Ferreira}\ \emph {et~al.}(1983)\citenamefont
  {Ferreira}, \citenamefont {King}, \citenamefont {Jaccarino}, \citenamefont
  {Cardy},\ and\ \citenamefont
  {Guggenheim}}]{Ferreira1983RandomFieldInducedDestructionOfThePhaseTransitionOfADiluted2DIsingAntiferromagnet}%
  \BibitemOpen
  \bibfield  {author} {\bibinfo {author} {\bibfnamefont {I.~B.}\ \bibnamefont
  {Ferreira}}, \bibinfo {author} {\bibfnamefont {A.~R.}\ \bibnamefont {King}},
  \bibinfo {author} {\bibfnamefont {V.}~\bibnamefont {Jaccarino}}, \bibinfo
  {author} {\bibfnamefont {J.~L.}\ \bibnamefont {Cardy}},\ and\ \bibinfo
  {author} {\bibfnamefont {H.~J.}\ \bibnamefont {Guggenheim}},\ }\bibfield
  {title} {\bibinfo {title} {Random-field-induced destruction of the phase
  transition of a diluted two-dimensional {I}sing antiferromagnet:
  {$\mathrm{Rb}_{2} \mathrm{Co}_{0.85} \mathrm{Mg}_{0.15} \mathrm{F}_{4}$ }},\
  }\href@noop {} {\bibfield  {journal} {\bibinfo  {journal} {Phys. Rev. B}\
  }\textbf {\bibinfo {volume} {{28}}},\ \bibinfo {pages} {5192} (\bibinfo
  {year} {1983})}\BibitemShut {NoStop}%
\bibitem [{\citenamefont
  {Griffiths}(1969)}]{Griffiths1969NonanalyticBehaviorAboveTheCriticalPointInARandomIsingFerromagnet}%
  \BibitemOpen
  \bibfield  {author} {\bibinfo {author} {\bibfnamefont {R.~B.}\ \bibnamefont
  {Griffiths}},\ }\bibfield  {title} {\bibinfo {title} {Nonanalytic behavior
  above the critical point in a random {I}sing ferromagnet},\ }\href@noop {}
  {\bibfield  {journal} {\bibinfo  {journal} {Phys. Rev. Lett.}\ }\textbf
  {\bibinfo {volume} {{23}}},\ \bibinfo {pages} {17} (\bibinfo {year}
  {1969})}\BibitemShut {NoStop}%
\bibitem [{\citenamefont {Bray}(1987)}]{Bray1987NatureOfTheGriffithsPhase}%
  \BibitemOpen
  \bibfield  {author} {\bibinfo {author} {\bibfnamefont {A.~J.}\ \bibnamefont
  {Bray}},\ }\bibfield  {title} {\bibinfo {title} {Nature of the {G}riffiths
  phase},\ }\href@noop {} {\bibfield  {journal} {\bibinfo  {journal} {Phys.
  Rev. Lett.}\ }\textbf {\bibinfo {volume} {{59}}},\ \bibinfo {pages} {586}
  (\bibinfo {year} {1987})}\BibitemShut {NoStop}%
\bibitem [{\citenamefont {Bray}\ and\ \citenamefont
  {Moore}(1982)}]{BrayMoore1982OnTheEigenvalueSpectrumOfTheSusceptibilityMatrixForRandomSpinSystems}%
  \BibitemOpen
  \bibfield  {author} {\bibinfo {author} {\bibfnamefont {A.~J.}\ \bibnamefont
  {Bray}}\ and\ \bibinfo {author} {\bibfnamefont {M.~A.}\ \bibnamefont
  {Moore}},\ }\bibfield  {title} {\bibinfo {title} {On the eigenvalue spectrum
  of the susceptibility matrix for random spin systems},\ }\href@noop {}
  {\bibfield  {journal} {\bibinfo  {journal} {J. Phys. C}\ }\textbf {\bibinfo
  {volume} {{15}}},\ \bibinfo {pages} {L765} (\bibinfo {year}
  {1982})}\BibitemShut {NoStop}%
\bibitem [{\citenamefont
  {Bray}(1988)}]{Bray1988DynamicsOfDiluteMagnetsAboveTc}%
  \BibitemOpen
  \bibfield  {author} {\bibinfo {author} {\bibfnamefont {A.~J.}\ \bibnamefont
  {Bray}},\ }\bibfield  {title} {\bibinfo {title} {Dynamics of dilute magnets
  above {$T_{\mathrm{c}}$}},\ }\href@noop {} {\bibfield  {journal} {\bibinfo
  {journal} {Phys. Rev. Lett.}\ }\textbf {\bibinfo {volume} {{60}}},\ \bibinfo
  {pages} {720} (\bibinfo {year} {1988})}\BibitemShut {NoStop}%
\bibitem [{\citenamefont {Colborne}\ and\ \citenamefont
  {Bray}(1989)}]{ColborneBray1989MonteCarloStudyOfGriffithsPhaseDynamicsInDiluteFerromagnets}%
  \BibitemOpen
  \bibfield  {author} {\bibinfo {author} {\bibfnamefont {S.}~\bibnamefont
  {Colborne}}\ and\ \bibinfo {author} {\bibfnamefont {A.}~\bibnamefont
  {Bray}},\ }\bibfield  {title} {\bibinfo {title} {{M}onte {C}arlo study of
  {G}riffiths phase dynamics in dilute ferromagnets},\ }\href@noop {}
  {\bibfield  {journal} {\bibinfo  {journal} {J. of Phys. A}\ }\textbf
  {\bibinfo {volume} {{22}}},\ \bibinfo {pages} {2505} (\bibinfo {year}
  {1989})}\BibitemShut {NoStop}%
\bibitem [{\citenamefont {Heng}\ \emph {et~al.}(2006)\citenamefont {Heng},
  \citenamefont {Song-Liu}, \citenamefont {Jing-Lin}, \citenamefont {Xiu-Lin},
  \citenamefont {Pai}, \citenamefont {Yong-Qiang},\ and\ \citenamefont
  {Li}}]{HengEtAl2006MonteCarloStudyOfGriffithsPhaseInRandomlySiteDilutedIsingMagneticSystem}%
  \BibitemOpen
  \bibfield  {author} {\bibinfo {author} {\bibfnamefont {C.}~\bibnamefont
  {Heng}}, \bibinfo {author} {\bibfnamefont {Y.}~\bibnamefont {Song-Liu}},
  \bibinfo {author} {\bibfnamefont {S.}~\bibnamefont {Jing-Lin}}, \bibinfo
  {author} {\bibfnamefont {J.}~\bibnamefont {Xiu-Lin}}, \bibinfo {author}
  {\bibfnamefont {L.}~\bibnamefont {Pai}}, \bibinfo {author} {\bibfnamefont
  {W.}~\bibnamefont {Yong-Qiang}},\ and\ \bibinfo {author} {\bibfnamefont
  {L.}~\bibnamefont {Li}},\ }\bibfield  {title} {\bibinfo {title} {{M}onte
  {C}arlo study of {G}riffiths phase in randomly site diluted {I}sing magnetic
  system},\ }\href@noop {} {\bibfield  {journal} {\bibinfo  {journal} {Chin.
  Phys. Lett.}\ }\textbf {\bibinfo {volume} {{23}}},\ \bibinfo {pages} {1176}
  (\bibinfo {year} {2006})}\BibitemShut {NoStop}%
\bibitem [{\citenamefont {Bray}\ and\ \citenamefont
  {Huifang}(1989)}]{BrayHuifang1989GriffithsSingularitiesInRandomMagnetsResultsForASolubleModel}%
  \BibitemOpen
  \bibfield  {author} {\bibinfo {author} {\bibfnamefont {A.~J.}\ \bibnamefont
  {Bray}}\ and\ \bibinfo {author} {\bibfnamefont {D.}~\bibnamefont {Huifang}},\
  }\bibfield  {title} {\bibinfo {title} {{G}riffiths singularities in random
  magnets: results for a soluble model},\ }\href@noop {} {\bibfield  {journal}
  {\bibinfo  {journal} {Phys. Rev. B}\ }\textbf {\bibinfo {volume} {{40}}},\
  \bibinfo {pages} {6980} (\bibinfo {year} {1989})}\BibitemShut {NoStop}%
\bibitem [{\citenamefont {Matsuda}\ \emph {et~al.}(2008)\citenamefont
  {Matsuda}, \citenamefont {Nishimori},\ and\ \citenamefont
  {Hukushima}}]{MatsudaEtAl2008TheDistributionOfLeeYangZerosAndGriffithsSingularitiesInThePMJModelOfSGs}%
  \BibitemOpen
  \bibfield  {author} {\bibinfo {author} {\bibfnamefont {Y.}~\bibnamefont
  {Matsuda}}, \bibinfo {author} {\bibfnamefont {H.}~\bibnamefont {Nishimori}},\
  and\ \bibinfo {author} {\bibfnamefont {K.}~\bibnamefont {Hukushima}},\
  }\bibfield  {title} {\bibinfo {title} {The distribution of {L}ee-{Y}ang zeros
  and {G}riffiths singularities in the {$\pm J$} model of spin glasses},\
  }\href@noop {} {\bibfield  {journal} {\bibinfo  {journal} {J. Phys. A}\
  }\textbf {\bibinfo {volume} {{41}}},\ \bibinfo {pages} {324012} (\bibinfo
  {year} {2008})}\BibitemShut {NoStop}%
\bibitem [{\citenamefont
  {Vojta}(2010)}]{Vojta2010QuantumGriffithsEffectsAndSmearedPhaseTransitionsInMetalsTheoryAndExperiment}%
  \BibitemOpen
  \bibfield  {author} {\bibinfo {author} {\bibfnamefont {T.}~\bibnamefont
  {Vojta}},\ }\bibfield  {title} {\bibinfo {title} {Quantum {G}riffiths effects
  and smeared phase transitions in metals: theory and experiment},\ }\href@noop
  {} {\bibfield  {journal} {\bibinfo  {journal} {Low Temp. Phys.}\ }\textbf
  {\bibinfo {volume} {{161}}},\ \bibinfo {pages} {299} (\bibinfo {year}
  {2010})}\BibitemShut {NoStop}%
\bibitem [{\citenamefont
  {Fisher}(1992)}]{Fisher1992RandomTransverseFieldIsingSpinChains}%
  \BibitemOpen
  \bibfield  {author} {\bibinfo {author} {\bibfnamefont {D.~S.}\ \bibnamefont
  {Fisher}},\ }\bibfield  {title} {\bibinfo {title} {Random transverse field
  {I}sing spin chains},\ }\href@noop {} {\bibfield  {journal} {\bibinfo
  {journal} {Phys. Rev. Lett.}\ }\textbf {\bibinfo {volume} {{69}}},\ \bibinfo
  {pages} {534} (\bibinfo {year} {1992})}\BibitemShut {NoStop}%
\bibitem [{\citenamefont
  {Fisher}(1995)}]{Fisher1995CriticalBehaviorOfRandomTransverseFieldIsingSpinChains}%
  \BibitemOpen
  \bibfield  {author} {\bibinfo {author} {\bibfnamefont {D.~S.}\ \bibnamefont
  {Fisher}},\ }\bibfield  {title} {\bibinfo {title} {Critical behavior of
  random transverse-field {I}sing spin chains},\ }\href@noop {} {\bibfield
  {journal} {\bibinfo  {journal} {Phys. Rev. B}\ }\textbf {\bibinfo {volume}
  {{51}}},\ \bibinfo {pages} {6411} (\bibinfo {year} {1995})}\BibitemShut
  {NoStop}%
\bibitem [{\citenamefont {Young}\ and\ \citenamefont
  {Rieger}(1996)}]{YoungRieger1996NumericalStudyOfTheRandomTransverseFieldIsingSpinChain}%
  \BibitemOpen
  \bibfield  {author} {\bibinfo {author} {\bibfnamefont {A.~P.}\ \bibnamefont
  {Young}}\ and\ \bibinfo {author} {\bibfnamefont {H.}~\bibnamefont {Rieger}},\
  }\bibfield  {title} {\bibinfo {title} {Numerical study of the random
  transverse-field {I}sing spin chain},\ }\href@noop {} {\bibfield  {journal}
  {\bibinfo  {journal} {Phys. Rev. B}\ }\textbf {\bibinfo {volume} {{53}}},\
  \bibinfo {pages} {8486} (\bibinfo {year} {1996})}\BibitemShut {NoStop}%
\bibitem [{\citenamefont {Pich}\ \emph {et~al.}(1998)\citenamefont {Pich},
  \citenamefont {Young}, \citenamefont {Rieger},\ and\ \citenamefont
  {Kawashima}}]{PichEtAl1998CriticalBehaviorAndGriffithsMcCoySingularitiesInThe2DRandomQuantumIsingFerromagnet}%
  \BibitemOpen
  \bibfield  {author} {\bibinfo {author} {\bibfnamefont {C.}~\bibnamefont
  {Pich}}, \bibinfo {author} {\bibfnamefont {A.~P.}\ \bibnamefont {Young}},
  \bibinfo {author} {\bibfnamefont {H.}~\bibnamefont {Rieger}},\ and\ \bibinfo
  {author} {\bibfnamefont {N.}~\bibnamefont {Kawashima}},\ }\bibfield  {title}
  {\bibinfo {title} {Critical behavior and {G}riffiths-{McCoy} singularities in
  the two-dimensional random quantum {I}sing ferromagnet},\ }\href@noop {}
  {\bibfield  {journal} {\bibinfo  {journal} {Phys. Rev. Lett.}\ }\textbf
  {\bibinfo {volume} {{81}}},\ \bibinfo {pages} {5916} (\bibinfo {year}
  {1998})}\BibitemShut {NoStop}%
\bibitem [{\citenamefont {Wang}\ \emph {et~al.}(2017)\citenamefont {Wang},
  \citenamefont {Gebretsadik}, \citenamefont {Ubaid-Kassis}, \citenamefont
  {Schroeder}, \citenamefont {Vojta}, \citenamefont {Baker}, \citenamefont
  {Pratt}, \citenamefont {Blundell}, \citenamefont {Lancaster}, \citenamefont
  {Franke}, \citenamefont {M\"oller},\ and\ \citenamefont
  {Page}}]{WangEtAl2017QuantumGriffithsPhaseExperiment}%
  \BibitemOpen
  \bibfield  {author} {\bibinfo {author} {\bibfnamefont {R.}~\bibnamefont
  {Wang}}, \bibinfo {author} {\bibfnamefont {A.}~\bibnamefont {Gebretsadik}},
  \bibinfo {author} {\bibfnamefont {S.}~\bibnamefont {Ubaid-Kassis}}, \bibinfo
  {author} {\bibfnamefont {A.}~\bibnamefont {Schroeder}}, \bibinfo {author}
  {\bibfnamefont {T.}~\bibnamefont {Vojta}}, \bibinfo {author} {\bibfnamefont
  {P.~J.}\ \bibnamefont {Baker}}, \bibinfo {author} {\bibfnamefont {F.~L.}\
  \bibnamefont {Pratt}}, \bibinfo {author} {\bibfnamefont {S.~J.}\ \bibnamefont
  {Blundell}}, \bibinfo {author} {\bibfnamefont {T.}~\bibnamefont {Lancaster}},
  \bibinfo {author} {\bibfnamefont {I.}~\bibnamefont {Franke}}, \bibinfo
  {author} {\bibfnamefont {J.~S.}\ \bibnamefont {M\"oller}},\ and\ \bibinfo
  {author} {\bibfnamefont {K.}~\bibnamefont {Page}},\ }\bibfield  {title}
  {\bibinfo {title} {Quantum {G}riffiths phase inside the ferromagnetic phase
  of {${\mathrm{Ni}}_{1\ensuremath{-}x}{\mathrm{V}}_{x}$}},\ }\href@noop {}
  {\bibfield  {journal} {\bibinfo  {journal} {Phys. Rev. Lett.}\ }\textbf
  {\bibinfo {volume} {{118}}},\ \bibinfo {pages} {267202} (\bibinfo {year}
  {2017})}\BibitemShut {NoStop}%
\bibitem [{\citenamefont {Nishimura}\ \emph {et~al.}(2020)\citenamefont
  {Nishimura}, \citenamefont {Nishimori},\ and\ \citenamefont
  {Katzgraber}}]{NishimuraEtAl2020GriffithsMcCoySingularityonTheDilutedChimeraGraph}%
  \BibitemOpen
  \bibfield  {author} {\bibinfo {author} {\bibfnamefont {K.}~\bibnamefont
  {Nishimura}}, \bibinfo {author} {\bibfnamefont {H.}~\bibnamefont
  {Nishimori}},\ and\ \bibinfo {author} {\bibfnamefont {H.~G.}\ \bibnamefont
  {Katzgraber}},\ }\bibfield  {title} {\bibinfo {title} {{G}riffiths-{McCoy}
  singularity on the diluted {C}himera graph: {M}onte {C}arlo simulations and
  experiments on quantum hardware},\ }\href@noop {} {\bibfield  {journal}
  {\bibinfo  {journal} {Phys. Rev. A}\ }\textbf {\bibinfo {volume} {{102}}},\
  \bibinfo {pages} {042403} (\bibinfo {year} {2020})}\BibitemShut {NoStop}%
\bibitem [{\citenamefont {Hartmann}(2015)}]{practical_guide2015}%
  \BibitemOpen
  \bibfield  {author} {\bibinfo {author} {\bibfnamefont {A.~K.}\ \bibnamefont
  {Hartmann}},\ }\href@noop {} {\emph {\bibinfo {title} {{Big Practical Guide
  to Computer Simulations}}}}\ (\bibinfo  {publisher} {World Scientific},\
  \bibinfo {address} {Singapore},\ \bibinfo {year} {2015})\BibitemShut
  {NoStop}%
\bibitem [{\citenamefont {Newman}\ and\ \citenamefont
  {Barkema}(1999)}]{NewmanBarkema1999MonteCarloMethodsInStatisticalPhysics}%
  \BibitemOpen
  \bibfield  {author} {\bibinfo {author} {\bibfnamefont {M.~E.}\ \bibnamefont
  {Newman}}\ and\ \bibinfo {author} {\bibfnamefont {G.~T.}\ \bibnamefont
  {Barkema}},\ }\href@noop {} {\emph {\bibinfo {title} {Monte Carlo methods in
  statistical physics}}}\ (\bibinfo  {publisher} {Oxford University Press},\
  \bibinfo {address} {Oxford},\ \bibinfo {year} {1999})\BibitemShut {NoStop}%
\bibitem [{\citenamefont
  {Hartmann}(2002)}]{Hartmann2002SamplingRareEventsStatisticsOfLocalSequenceAlignments}%
  \BibitemOpen
  \bibfield  {author} {\bibinfo {author} {\bibfnamefont {A.~K.}\ \bibnamefont
  {Hartmann}},\ }\bibfield  {title} {\bibinfo {title} {Sampling rare events:
  statistics of local sequence alignments},\ }\href@noop {} {\bibfield
  {journal} {\bibinfo  {journal} {Phys. Rev. E}\ }\textbf {\bibinfo {volume}
  {{65}}},\ \bibinfo {pages} {056102} (\bibinfo {year} {2002})}\BibitemShut
  {NoStop}%
\bibitem [{\citenamefont
  {Hartmann}(2011)}]{Hartmann2011LargeDeviationPropertiesOfLargestComponentForRandomGraphs}%
  \BibitemOpen
  \bibfield  {author} {\bibinfo {author} {\bibfnamefont {A.~K.}\ \bibnamefont
  {Hartmann}},\ }\bibfield  {title} {\bibinfo {title} {Large-deviation
  properties of largest component for random graphs},\ }\href@noop {}
  {\bibfield  {journal} {\bibinfo  {journal} {Eur. Phys. J. B}\ }\textbf
  {\bibinfo {volume} {{84}}},\ \bibinfo {pages} {627} (\bibinfo {year}
  {2011})}\BibitemShut {NoStop}%
\bibitem [{\citenamefont {Hukushima}\ and\ \citenamefont
  {Iba}(2008)}]{HukushimaIba2008AMonteCarloAlgorithmForSamplingRareEvents}%
  \BibitemOpen
  \bibfield  {author} {\bibinfo {author} {\bibfnamefont {K.}~\bibnamefont
  {Hukushima}}\ and\ \bibinfo {author} {\bibfnamefont {Y.}~\bibnamefont
  {Iba}},\ }\bibfield  {title} {\bibinfo {title} {A {M}onte {C}arlo algorithm
  for sampling rare events: application to a search for the {G}riffiths
  singularity},\ }\href@noop {} {\bibfield  {journal} {\bibinfo  {journal} {J.
  Phys. Conf. Ser.}\ }\textbf {\bibinfo {volume} {{95}}},\ \bibinfo {pages}
  {012005} (\bibinfo {year} {2008})}\BibitemShut {NoStop}%
\bibitem [{\citenamefont {Neuhaus}\ and\ \citenamefont
  {Hager}(2006)}]{NeuhausHager2006FreeEnergyCalculationsWithMultipleGaussianModifiedEnsembles}%
  \BibitemOpen
  \bibfield  {author} {\bibinfo {author} {\bibfnamefont {T.}~\bibnamefont
  {Neuhaus}}\ and\ \bibinfo {author} {\bibfnamefont {J.~S.}\ \bibnamefont
  {Hager}},\ }\bibfield  {title} {\bibinfo {title} {Free-energy calculations
  with multiple {G}aussian modified ensembles},\ }\href@noop {} {\bibfield
  {journal} {\bibinfo  {journal} {Phys. Rev. E}\ }\textbf {\bibinfo {volume}
  {{74}}},\ \bibinfo {pages} {036702} (\bibinfo {year} {2006})}\BibitemShut
  {NoStop}%
\bibitem [{\citenamefont
  {Ohzeki}(2009)}]{Ohzeki2009LocationsOfMulticriticalPointsForSpinGlassesOnRegularLattices}%
  \BibitemOpen
  \bibfield  {author} {\bibinfo {author} {\bibfnamefont {M.}~\bibnamefont
  {Ohzeki}},\ }\bibfield  {title} {\bibinfo {title} {Locations of multicritical
  points for spin glasses on regular lattices},\ }\href@noop {} {\bibfield
  {journal} {\bibinfo  {journal} {Phys. Rev. E}\ }\textbf {\bibinfo {volume}
  {{79}}},\ \bibinfo {pages} {021129} (\bibinfo {year} {2009})}\BibitemShut
  {NoStop}%
\bibitem [{Note1()}]{Note1}%
  \BibitemOpen
  \bibinfo {note} {Note that below the ferromagnetic phase transition the
  symmetry is broken such that there are two thermodynamic states, one with
  positive and one with negative magnetization. To obtain the correct value for
  the magnetization, the average has to be restricted to one of the states. In
  numerical simulations, however, there is usually no symmetry breaking and
  thus one often uses the absolute value of the magnetization as an estimator
  for the magnetization or, alternatively, the largest FKCK cluster, see
  Sec.~\ref {sec:large_deviations_sampling}.}\BibitemShut {Stop}%
\bibitem [{\citenamefont
  {Nishimori}(1979)}]{Nishimori1979ConjectureOnTheExactTransitionPointOfTheRandomIsingFerromagnet}%
  \BibitemOpen
  \bibfield  {author} {\bibinfo {author} {\bibfnamefont {H.}~\bibnamefont
  {Nishimori}},\ }\bibfield  {title} {\bibinfo {title} {Conjecture on the exact
  transition point of the random {I}sing ferromagnet},\ }\href@noop {}
  {\bibfield  {journal} {\bibinfo  {journal} {J. Phys. C}\ }\textbf {\bibinfo
  {volume} {{12}}},\ \bibinfo {pages} {L905} (\bibinfo {year}
  {1979})}\BibitemShut {NoStop}%
\bibitem [{\citenamefont {Zhong}\ \emph {et~al.}(2020)\citenamefont {Zhong},
  \citenamefont {Barkema},\ and\ \citenamefont
  {Panja}}]{ZhongEtAl2020SuperSlowingDownInTheBondDilutedIsingModel}%
  \BibitemOpen
  \bibfield  {author} {\bibinfo {author} {\bibfnamefont {W.}~\bibnamefont
  {Zhong}}, \bibinfo {author} {\bibfnamefont {G.~T.}\ \bibnamefont {Barkema}},\
  and\ \bibinfo {author} {\bibfnamefont {D.}~\bibnamefont {Panja}},\ }\bibfield
   {title} {\bibinfo {title} {Super slowing down in the bond-diluted {I}sing
  model},\ }\href@noop {} {\bibfield  {journal} {\bibinfo  {journal} {Phys.
  Rev. E}\ }\textbf {\bibinfo {volume} {{102}}},\ \bibinfo {pages} {022132}
  (\bibinfo {year} {2020})}\BibitemShut {NoStop}%
\bibitem [{Note2()}]{Note2}%
  \BibitemOpen
  \bibinfo {note} {Note that in most cases we used $Y_{\protect \bm
  {J}}=\protect \sqrt {N \chi _{\protect \bm {J}}}$ instead of $\chi _{\protect
  \bm {J}}$ since this quantity is found to be easier to sample. The
  distribution of $\chi _{\protect \bm {J}}$ can then be obtained by a change
  of variables.}\BibitemShut {Stop}%
\bibitem [{\citenamefont
  {Schawe}(2019)}]{Schawe2019LargeDeviationsOfConvexHullsETC}%
  \BibitemOpen
  \bibfield  {author} {\bibinfo {author} {\bibfnamefont {H.}~\bibnamefont
  {Schawe}},\ }\emph {\bibinfo {title} {Large Deviations of Convex Hulls of
  Random Walks and Other Stochastic Models}},\ \href@noop {} {\bibinfo {type}
  {dissertation}},\ \bibinfo  {school} {Carl von Ossietzky Universit{\"a}t
  Oldenburg}, \bibinfo {address} {Oldenburg} (\bibinfo {year}
  {2019})\BibitemShut {NoStop}%
\bibitem [{\citenamefont {Ebert}\ \emph {et~al.}(2022)\citenamefont {Ebert},
  \citenamefont {Gessert},\ and\ \citenamefont {Weigel}}]{ebert:22}%
  \BibitemOpen
  \bibfield  {author} {\bibinfo {author} {\bibfnamefont {P.~L.}\ \bibnamefont
  {Ebert}}, \bibinfo {author} {\bibfnamefont {D.}~\bibnamefont {Gessert}},\
  and\ \bibinfo {author} {\bibfnamefont {M.}~\bibnamefont {Weigel}},\
  }\bibfield  {title} {\bibinfo {title} {Weighted averages in population
  annealing: analysis and general framework},\ }\href
  {https://doi.org/10.1103/PhysRevE.106.045303} {\bibfield  {journal} {\bibinfo
   {journal} {Phys. Rev. E}\ }\textbf {\bibinfo {volume} {106}},\ \bibinfo
  {pages} {045303} (\bibinfo {year} {2022})}\BibitemShut {NoStop}%
\bibitem [{\citenamefont {Swendsen}\ and\ \citenamefont
  {Wang}(1987)}]{SwendsenWang1987NonuniversalCriticalDynamicsInMonteCarloSimulations}%
  \BibitemOpen
  \bibfield  {author} {\bibinfo {author} {\bibfnamefont {R.~H.}\ \bibnamefont
  {Swendsen}}\ and\ \bibinfo {author} {\bibfnamefont {J.~S.}\ \bibnamefont
  {Wang}},\ }\bibfield  {title} {\bibinfo {title} {Nonuniversal critical
  dynamics in monte carlo simulations},\ }\href@noop {} {\bibfield  {journal}
  {\bibinfo  {journal} {Phys. Rev. Lett.}\ }\textbf {\bibinfo {volume}
  {{58}}},\ \bibinfo {pages} {86} (\bibinfo {year} {1987})}\BibitemShut
  {NoStop}%
\bibitem [{\citenamefont {Kole}\ \emph {et~al.}(2022)\citenamefont {Kole},
  \citenamefont {Barkema},\ and\ \citenamefont
  {Fritz}}]{Kole2022ComparisonOfClusterAlgorithmsForTheBondDilutedIsingModel}%
  \BibitemOpen
  \bibfield  {author} {\bibinfo {author} {\bibfnamefont {A.~H.}\ \bibnamefont
  {Kole}}, \bibinfo {author} {\bibfnamefont {G.~T.}\ \bibnamefont {Barkema}},\
  and\ \bibinfo {author} {\bibfnamefont {L.}~\bibnamefont {Fritz}},\ }\bibfield
   {title} {\bibinfo {title} {Comparison of cluster algorithms for the
  bond-diluted {I}sing model},\ }\href@noop {} {\bibfield  {journal} {\bibinfo
  {journal} {Phys. Rev. E}\ }\textbf {\bibinfo {volume} {{105}}},\ \bibinfo
  {pages} {015313} (\bibinfo {year} {2022})}\BibitemShut {NoStop}%
\bibitem [{\citenamefont {Wolff}(1989)}]{wolff:89}%
  \BibitemOpen
  \bibfield  {author} {\bibinfo {author} {\bibfnamefont {U.}~\bibnamefont
  {Wolff}},\ }\bibfield  {title} {\bibinfo {title} {Comparison between cluster
  {Monte Carlo} algorithms in the {Ising} model},\ }\href
  {https://doi.org/10.1016/0370-2693(89)91563-3} {\bibfield  {journal}
  {\bibinfo  {journal} {Phys. Lett. B}\ }\textbf {\bibinfo {volume} {228}},\
  \bibinfo {pages} {379} (\bibinfo {year} {1989})}\BibitemShut {NoStop}%
\bibitem [{Note3()}]{Note3}%
  \BibitemOpen
  \bibinfo {note} {The FKCK clusters are also called FK clusters or CK droplets
  due to their historical origin \cite
  {ConiglioFierro2021CorrelatedPercolation}, see also Ref.~\cite
  {munster:23}.}\BibitemShut {Stop}%
\bibitem [{\citenamefont {Binder}\ and\ \citenamefont
  {Heermann}(2010)}]{BinderHeermann2010SomeImportantRecentDevelopmentsOfTheMCMethology}%
  \BibitemOpen
  \bibfield  {author} {\bibinfo {author} {\bibfnamefont {K.}~\bibnamefont
  {Binder}}\ and\ \bibinfo {author} {\bibfnamefont {D.~W.}\ \bibnamefont
  {Heermann}},\ }\bibinfo {title} {Some important recent developments of the
  {M}onte {C}arlo methodology},\ in\ \href@noop {} {\emph {\bibinfo {booktitle}
  {{M}onte {C}arlo Simulation in Statistical Physics: An Introduction}}}\
  (\bibinfo  {publisher} {Springer Berlin Heidelberg},\ \bibinfo {address}
  {Berlin, Heidelberg},\ \bibinfo {year} {2010})\ p.\ \bibinfo {pages}
  {111}\BibitemShut {NoStop}%
\bibitem [{\citenamefont {Roussenq}\ \emph {et~al.}(1982)\citenamefont
  {Roussenq}, \citenamefont {Coniglio},\ and\ \citenamefont
  {Stauffer}}]{RoussenqConiglioStauffer1982StudyOfDropletsForCorrelatedSiteBondPercolationIn3D}%
  \BibitemOpen
  \bibfield  {author} {\bibinfo {author} {\bibfnamefont {J.}~\bibnamefont
  {Roussenq}}, \bibinfo {author} {\bibfnamefont {A.}~\bibnamefont {Coniglio}},\
  and\ \bibinfo {author} {\bibfnamefont {D.}~\bibnamefont {Stauffer}},\
  }\bibfield  {title} {\bibinfo {title} {Study of droplets for correlated
  site-bond percolation in three dimensions},\ }\href@noop {} {\bibfield
  {journal} {\bibinfo  {journal} {Journal de Physique Lettres}\ }\textbf
  {\bibinfo {volume} {{43}}},\ \bibinfo {pages} {703} (\bibinfo {year}
  {1982})}\BibitemShut {NoStop}%
\bibitem [{\citenamefont {Weigel}\ \emph {et~al.}(2002)\citenamefont {Weigel},
  \citenamefont {Janke},\ and\ \citenamefont {Hu}}]{weigel:02a}%
  \BibitemOpen
  \bibfield  {author} {\bibinfo {author} {\bibfnamefont {M.}~\bibnamefont
  {Weigel}}, \bibinfo {author} {\bibfnamefont {W.}~\bibnamefont {Janke}},\ and\
  \bibinfo {author} {\bibfnamefont {C.~K.}\ \bibnamefont {Hu}},\ }\bibfield
  {title} {\bibinfo {title} {Random-cluster multihistogram sampling for the
  \(q\)-state potts model},\ }\href@noop {} {\bibfield  {journal} {\bibinfo
  {journal} {Phys. Rev. E}\ }\textbf {\bibinfo {volume} {65}},\ \bibinfo
  {pages} {036109} (\bibinfo {year} {2002})}\BibitemShut {NoStop}%
\bibitem [{\citenamefont
  {Young}(2015)}]{Young2015EverythingYouWantedToKnowAboutDataAnalysis}%
  \BibitemOpen
  \bibfield  {author} {\bibinfo {author} {\bibfnamefont {A.~P.}\ \bibnamefont
  {Young}},\ }\href@noop {} {\emph {\bibinfo {title} {{Everything You Wanted to
  Know About Data Analysis and Fitting but Were Afraid to Ask}}}}\ (\bibinfo
  {publisher} {Springer},\ \bibinfo {address} {Cham},\ \bibinfo {year}
  {2015})\BibitemShut {NoStop}%
\bibitem [{\citenamefont
  {Touchette}(2009)}]{Touchette2009TheLargeDeviationApproachToStatisticalMechanics}%
  \BibitemOpen
  \bibfield  {author} {\bibinfo {author} {\bibfnamefont {H.}~\bibnamefont
  {Touchette}},\ }\bibfield  {title} {\bibinfo {title} {The large deviation
  approach to statistical mechanics},\ }\href@noop {} {\bibfield  {journal}
  {\bibinfo  {journal} {Phys. Rep.}\ }\textbf {\bibinfo {volume} {478}}
  (\bibinfo {year} {2009})}\BibitemShut {NoStop}%
\bibitem [{\citenamefont {Wiseman}\ and\ \citenamefont
  {Domany}(1995)}]{WisemanDomany1995LackOfSelfAveragingInCriticalDisorderedSystems}%
  \BibitemOpen
  \bibfield  {author} {\bibinfo {author} {\bibfnamefont {S.}~\bibnamefont
  {Wiseman}}\ and\ \bibinfo {author} {\bibfnamefont {E.}~\bibnamefont
  {Domany}},\ }\bibfield  {title} {\bibinfo {title} {Lack of self-averaging in
  critical disordered systems},\ }\href@noop {} {\bibfield  {journal} {\bibinfo
   {journal} {Phys. Rev. E}\ }\textbf {\bibinfo {volume} {{52}}},\ \bibinfo
  {pages} {3469} (\bibinfo {year} {1995})}\BibitemShut {NoStop}%
\bibitem [{\citenamefont {Wiseman}\ and\ \citenamefont
  {Domany}(1998)}]{WisemanDomany1998FSSAndLackOfSelfAveragingInCriticalDisorderedSystems}%
  \BibitemOpen
  \bibfield  {author} {\bibinfo {author} {\bibfnamefont {S.}~\bibnamefont
  {Wiseman}}\ and\ \bibinfo {author} {\bibfnamefont {E.}~\bibnamefont
  {Domany}},\ }\bibfield  {title} {\bibinfo {title} {Finite-size scaling and
  lack of self-averaging in critical disordered systems},\ }\href@noop {}
  {\bibfield  {journal} {\bibinfo  {journal} {Phys. Rev. Lett.}\ }\textbf
  {\bibinfo {volume} {81}},\ \bibinfo {pages} {22} (\bibinfo {year}
  {1998})}\BibitemShut {NoStop}%
\bibitem [{\citenamefont {Aharony}\ and\ \citenamefont
  {Harris}(1996)}]{AharonyHarris1996AbsenceOfSelfAveragingAndUniversalFluctuationsInRandomSystemsNearCriticalPoints}%
  \BibitemOpen
  \bibfield  {author} {\bibinfo {author} {\bibfnamefont {A.}~\bibnamefont
  {Aharony}}\ and\ \bibinfo {author} {\bibfnamefont {A.~B.}\ \bibnamefont
  {Harris}},\ }\bibfield  {title} {\bibinfo {title} {Absence of self-averaging
  and universal fluctuations in random systems near critical points},\
  }\href@noop {} {\bibfield  {journal} {\bibinfo  {journal} {Phys. Rev. Lett.}\
  }\textbf {\bibinfo {volume} {{77}}},\ \bibinfo {pages} {3700} (\bibinfo
  {year} {1996})}\BibitemShut {NoStop}%
\bibitem [{Note4()}]{Note4}%
  \BibitemOpen
  \bibinfo {note} {As a technical detail we want to mention that for the plots
  in Fig.~\ref {fig:df_below_tc_distr_plot}, Fig.~\ref
  {fig:df_beta1x5_distr_correlation_map_subplot} and Fig.~\ref
  {fig:df_ferromagnetic_bond_correlation_map_plot} we have used the common
  estimator $\chi _{\protect \bm {J}}=N (\langle \protect \hat {m}^2 \rangle _S
  -\langle \vert \protect \hat {m} \vert \rangle _S^2 )$ to compute the
  magnetic susceptibility since this worked better for sampling at larger
  system sizes. The results, however, are fully consistent with those of the
  cluster estimator of Eq.~\protect \textup {\hbox {\mathsurround \z@ \protect
  \normalfont (\ignorespaces \ref {eq:cluster_estimator}\unskip \@@italiccorr
  )}} for $T<T_{\protect \mathrm {c}}$}\BibitemShut {NoStop}%
\bibitem [{\citenamefont
  {Fisch}(1978)}]{Fisch1978CriticalTemperatureForTwoDimensionalIsingFerromagnetsWithQuenchedBondDisorder}%
  \BibitemOpen
  \bibfield  {author} {\bibinfo {author} {\bibfnamefont {R.}~\bibnamefont
  {Fisch}},\ }\bibfield  {title} {\bibinfo {title} {Critical temperature for
  two-dimensional {I}sing ferromagnets with quenched bond disorder},\
  }\href@noop {} {\bibfield  {journal} {\bibinfo  {journal} {J. Stat. Phys.}\
  }\textbf {\bibinfo {volume} {{18}}},\ \bibinfo {pages} {111} (\bibinfo {year}
  {1978})}\BibitemShut {NoStop}%
\bibitem [{Note5()}]{Note5}%
  \BibitemOpen
  \bibinfo {note} {In case of the random ferromagnet with $J_{\protect \bm {x}
  \protect \bm {y}}=\{c,1\}$ and $0 < c \leq 1$ the variance of the largest
  FKCK cluster might lead to a tail in the distribution of the magnetic
  susceptibility inside the ferromagnetic phase above zero temperature, but
  whether this is indeed the case should be investigated in future
  studies.}\BibitemShut {Stop}%
\bibitem [{\citenamefont {Melchert}(2009)}]{Melchert2009Autoscale}%
  \BibitemOpen
  \bibfield  {author} {\bibinfo {author} {\bibfnamefont {O.}~\bibnamefont
  {Melchert}},\ }\href@noop {} {\bibinfo {title} {{autoScale.py} - a program
  for automatic finite-size scaling analyses: a user's guide}} (\bibinfo {year}
  {2009}),\ \Eprint {https://arxiv.org/abs/0910.5403} {arXiv:0910.5403}
  \BibitemShut {NoStop}%
\bibitem [{\citenamefont {Malakis}\ and\ \citenamefont
  {Fytas}(2006)}]{MalakisFytas2006LackOfSelfAveragingOfTheSpecificHeatInTheThreeDimensionalRandomFieldIsingModel}%
  \BibitemOpen
  \bibfield  {author} {\bibinfo {author} {\bibfnamefont {A.}~\bibnamefont
  {Malakis}}\ and\ \bibinfo {author} {\bibfnamefont {N.~G.}\ \bibnamefont
  {Fytas}},\ }\bibfield  {title} {\bibinfo {title} {Lack of self-averaging of
  the specific heat in the three-dimensional random-field {I}sing model},\
  }\href@noop {} {\bibfield  {journal} {\bibinfo  {journal} {Phys. Rev. E}\
  }\textbf {\bibinfo {volume} {{73}}},\ \bibinfo {pages} {016109} (\bibinfo
  {year} {2006})}\BibitemShut {NoStop}%
\bibitem [{\citenamefont {Stauffer}\ and\ \citenamefont
  {Aharony}(1994)}]{StaufferAharony1994IntroductionToPercolationTheory}%
  \BibitemOpen
  \bibfield  {author} {\bibinfo {author} {\bibfnamefont {D.}~\bibnamefont
  {Stauffer}}\ and\ \bibinfo {author} {\bibfnamefont {A.}~\bibnamefont
  {Aharony}},\ }\href@noop {} {\emph {\bibinfo {title} {Introduction to
  Percolation Theory}}},\ \bibinfo {edition} {2nd}\ ed.\ (\bibinfo  {publisher}
  {Taylor \& Francis},\ \bibinfo {address} {London},\ \bibinfo {year}
  {1994})\BibitemShut {NoStop}%
\bibitem [{\citenamefont {M{\"u}nster}\ and\ \citenamefont
  {Weigel}(2023)}]{munster:23}%
  \BibitemOpen
  \bibfield  {author} {\bibinfo {author} {\bibfnamefont {L.}~\bibnamefont
  {M{\"u}nster}}\ and\ \bibinfo {author} {\bibfnamefont {M.}~\bibnamefont
  {Weigel}},\ }\bibfield  {title} {\bibinfo {title} {Cluster percolation in the
  two-dimensional ising spin glass},\ }\href@noop {} {\bibfield  {journal}
  {\bibinfo  {journal} {Phys. Rev. E}\ }\textbf {\bibinfo {volume} {107}},\
  \bibinfo {pages} {054103} (\bibinfo {year} {2023})}\BibitemShut {NoStop}%
\end{thebibliography}%
\end{document}